%% file: main.tex
\pdfoutput=1
\documentclass[12pt]{article}

\usepackage[fleqn]{amsmath}
\usepackage[T1]{fontenc}
\usepackage{times}
\usepackage{lmodern}
\usepackage{amssymb}
\usepackage{bm}
\usepackage{graphicx}
\usepackage{gensymb}
\usepackage{mymacros}
\usepackage{mathtools}
\usepackage{float}
\usepackage{dsfont}
\usepackage{geometry}
\usepackage{algpseudocode}
\usepackage{verbatim}
\usepackage{amsthm}
\usepackage{lineno}
\usepackage{caption}
\usepackage{subfig}
\usepackage{authblk}
\DeclareMathOperator{\Tr}{Tr}
\begin{document}

%\begin{frontmatter}


\title{Topology-faithful nonparametric estimation and tracking of bulk interface networks}


 \author[1]{S. Maddali\thanks{Corresponding author: smaddali@andrew.cmu.edu}}
 \author[2]{S. Ta'asan}
 \author[1,3]{R. M. Suter}

 \affil[1]{Dept. of Physics, Carnegie Mellon University, Pittsburgh PA 15213 (USA)}
 \affil[2]{Dept. of Mathematical Sciences, Carnegie Mellon University, Pittsburgh PA 15217 (USA)}
 \affil[3]{Dept. of Materials Science and Engineering, Carnegie Mellon University, Pittsburgh PA 15213 (USA)}

 \date{}

	\maketitle
%\author{S. Maddali, R. M. Suter}
%\address{Dept. of physics, Carnegie Mellon University, Pittsburgh, PA (USA)}

\begin{abstract}
\input{abstract.tex}
\end{abstract}

%\begin{keyword}
%Interface networks \sep Hierarchical smoothing \sep Non-parametric \sep Microstructure \sep Filtering \sep Grain boundary topology
%%% keywords here, in the form: keyword \sep keyword
%
%%% MSC codes here, in the form: \MSC code \sep code
%%% or \MSC[2008] code \sep code (2000 is the default)
%
%\end{keyword}

%\end{frontmatter}

%\linenumbers

\section{Introduction}
\label{S:intro}
\input{introduction.tex}

\section{General formalism}
\label{S:generalformalism}
\input{basicformalism.tex}

\section{Constrained smoothing}
\label{S:constrainedsmoothing}
\input{constrainedsmoothing.tex}

\section{$SMOOTH$ing a digitized curve in a plane}
\label{S:digitalplanarcurve}
\input{smoothingdigitizedcurves.tex}

\section{$SMOOTH$ing known shapes}
\label{S:knownshapes}
\input{smoothingknownshapes.tex}

%\section{Normal estimation and meshes}
%\label{sec:Errors}
%\input{errors.tex}

\section{Results: two-dimensional microstructure smoothing}
\label{S:2dmic}
\input{twodimmicsmooth.tex}

\section{Results: three-dimensional microstructure smoothing}
\label{S:3dmic}
\input{threedimmicsmooth.tex}

\section{Mesh quality}
\label{S:meshquality}
\input{meshquality.tex}

\section{Front tracking}
\label{S:fronttracking}
\input{fronttracking.tex}

\section{Summary and discussion}
\label{S:conclusion}
\input{conclusion.tex}

\section{Acknowledgements}
\label{S:ack}
\input{acknowledgements.tex}

%\newpage
\section{References}
%\bibliographystyle{unsrt}
%\bibliography{myrefs_smoothing}

\input{main.bbl}
\end{document}

%% file: abstract.tex
The main focus of this paper is a nonparametric filtering technique for the estimation of interface geometry in bulk materials obtainable from modern imaging measurements. 
The filtering methodology relies on an assumed hierarchy of topological features present in a typical interface network, such as foam interfaces and grain boundary networks in polycrystalline materials. 
Each type of topological feature is treated in order of rank in the hierarchy, with the lower-level feature being filtered subject to the positional constraints imposed by the higher-level features. 
Such a scheme is an alternative to existing surface smoothing/estimation techniques in microstructural materials science, in which the explicit treatment of different elements of the network topology is absent, or at best arbitrarily parameterized. 
We describe the ramifications of this technique in the usual microstructural applications in which the computation of important physical quantities is predicated on the precise estimation of the interface features. 
As an additional application, we describe a novel front-tracking algorithm for quantifying the transport of such interfaces.

%% file: introduction.tex
The morphology of surfaces and interfaces has garnered great interest in many fields of scientific and engineering research. Such studies have implications in applied physics, materials science, biology, pharmacology, chemical engineering and computer vision~\cite{McCready1986,McInerney1995,Lyklema2005}. A vast part of this research is predicated on the proper imaging of interfaces in the medium of interest. 
\input{bobsgiantedit.tex}

Whether from a basic or applied science viewpoint, the importance of characterizing grain boundaries in this manner cannot be overstated. In polycrystalline materials, the local interfacial energy density and mobility are known to be sensitive to the five grain boundary parameters at each location~\cite{Olmsted2009a,Olmsted2009,Ratanaphan2015,Bulatov2014}. It also informs applications like grain boundary engineering whose eventual goal is to precisely manipulate bulk material properties through the tuning of the grain boundary character distributions~\cite{Saylor2003,Saylor2003a,Saylor2004,Li2009}. Further, it is well-known that the topological elements of a grain boundary network like triple junctions and quad points are hotbeds of activity with respect to precipitate diffusion~\cite{King2010,Bokstein2001,Swiler1997} and strain accumulation~\cite{Rollett2010,Carter2012}. Real grain boundary networks are usually the starting point for atomistic and continuum simulations of microstructure evolution, the physics of which is most difficult to model at triple lines and quad points. 

All these applications are predicated upon measurements of the various topological features of a grain boundary network, which are inevitably subject to noise, whether through experimental resolution or image gridding. Given the generally accepted assumption of mesoscopically smooth interfaces, this necessitates the use of a smoothing estimator prior to any further analysis. Owing to the diverse roles of topological elements such as triple lines and quad points in microstructure phenomenology, an important motivation for this novel filtering technique and other recent ones~\cite{lee2014} is to give them their due importance through explicit treatment.

Other factors motivating this work are:
\begin{itemize}
	\item	Unlike voxelized images of most everyday objects, there exists no general intuition for the shape of a grain in a sample, and therefore a grain boundary. In the former case, iterative smoothing algorithms such as Laplace and Taubin smoothing~\cite{Taubin1995} yield an acceptable result that is partially helped along by the user's advance knowledge of the object in question. However these methods can suffer from under- or over-smoothing if the number of iterations or step size are not chosen properly. 
	\item	Explicit modeling techniques~\cite{Wang2009} more often than not belie the sheer variety in the observed structure of grain boundaries and network topologies. 
	\item	Existing nonparametric techniques~\cite{Lieberman2015} require the use of a smoothing window of a user-defined size. 
\end{itemize}
	The methodology described here internally optimizes a compromise between fidelity to the input data points and a constrained Laplacian smoothing. An objective function is minimized with respect to this compromise. The algorithm requires no user input in terms of filter parameters, only that the connectivity of the nodes be specified in advance, in the form of a graph. We distinguish the type of kernel resulting from graph-connectedness to a given node from a fixed-size window centered on that node since the former, which we rely upon, does not take into consideration the physical distance between neighboring nodes, and only keeps track of the connectivity. 

\input{bobssecondedit.tex}

\input{generalizedtopology.tex}

We first describe the topological hierarchy in general terms and then address the interface estimation procedure, which is a modification of Laplacian smoothing of a set of meshed surface points. This is followed by the application of the estimation algorithm to pixelated versions of easily parameterized geometric primitives, in particular circles, spheres and cylinders. Post-smoothing errors are quantified in terms of estimated sizes of these primitives as well as estimated normals for specific geometries. We then address specific cases of interest in mesoscale materials science: two- and three-dimensional grain boundary networks. The former finds relevance in the study of thin films and the latter in that of bulk material behavior (most prominently in the computation of grain boundary character distribution plots, a common characterization of materials microstructure). We demonstrate how the user is freed from the largely intuitive choices of smoothing parameters that is characteristic of iterative or windowed techniques. 
%Maximizing the extent of automation in these things is of great importance to software pipelines that streamline the processing and analysis of digital microstructure data~\cite{Groeber2014}. 
We then describe in some detail the applicability of this surface estimation algorithm to finite element methods in materials science as well as interface velocity estimation, which is a new capability made possible with data obtained from modern non-destructive imaging techniques. A new nonparameteric algorithm to achieve the latter is described.

%% file: bobsgiantedit.tex
Interfacial networks are composed of two-dimensional, possibly curved interfaces that separate two distinct regions of homogeneous matter, such as gas in bubble foams, or phases or crystalline orientations in solids. 
We use in this paper language relevant to interfaces in polycrystalline materials but the methods described are equally applicable in other fields by straightforward adaptation of the terminology. 
The three dimensional entities with more or less uniform crystalline characteristics henceforth will be referred to as `grains'. 

A particular type of grain boundary can be specified by five parameters on the mesoscale where `meso-' refers to a length scale that is large compared to interatomic distances but small compared to a typical grain size. Among the several possible parameterizations; we choose the set of three specifying the relative crystal orientations of the grains, and two specifying the local normal direction relative to the crystal axes in one of the grains. The normal direction in the other crystal frame can be computed from these five parameters. This parameterization ignores a microscopic relative translation on the atomic scale and thereby atomic-level faceting of the interface, a feature addressed explicitly in molecular statics and dynamics simulations. The set of these five parameters is said to specify the grain boundary character\cite{Saylor2004,Khorashadizadeh2011,Ratanaphan2014,Ratanaphan2015}. Note that the character between two grains can vary over the two-dimensional boundary between them because, while the misorientation is fixed, the local normal typically varies significantly over a curved grain surface. Similar characterizations can be made for triple lines (two misorientations and a tangent line) and quad points (three misorientations). Finally, we note that crystal symmetry is typically exploited to reduce these specifications to unique `fundamental zones' that span physically distinct ranges of orientations or misorientations.

%% file: bobssecondedit.tex
While the grains in polycrystals can take on essentially arbitrary shapes, the topological features typically encountered, and which are explicitly dealt with here, can be demonstrated with children's (or adult's) building blocks. Place two, say, cubic blocks (grains) together on a surface with edges aligned (blocks 1 and 2). The two blocks meet at a two dimensional interface (grain boundary 1-2). Now place a third block on the same surface so that it forms boundaries with both blocks 1 and 2 (boundaries 1-3 and 2-3). These boundaries meet at triple line 1-2-3. Now, place block 4 on top of these three so as to form boundaries with all three below (boundaries 1-4, 2-4, and 3-4). One now has new triple lines 1-2-4, 1-3-4, and 2-3-4. Furthermore, triple line 1-2-3 now terminates at a quadruple point, 1-2-3-4, where all four grains meet. Unless one makes special alignments to again align edges, these are the topological features that will characterize an extended group of similarly stacked blocks.

%% file: generalizedtopology.tex
In this paper, all line junctions of interfaces in a network are referred to by the generic term `triple line' in allusion to the fact that energetically stable junctions in 3D are shared between exactly three interfaces. The incidental existence of a `$n$-tuple line' in a polycrystalline material where $n > 3$ interfaces intersect is known to be energetically unstable, forcing the interface topology to deform to a lower energy configuration~\cite{Barrales-Mora2008,Gottstein2011}. Likewise, a node of termination of $n'$ such triple lines is referred to as a `quad point' irrespective of the value of $n'$, alluding to the fact that $n' = 4$ is the physically stable configuration in bulk materials. The mathematical machinery developed in this paper is as appropriate for contrived interface networks that deviate from this this topological rule as for digital images of real bulk microstructure, in which these deviations are almost never observed.

%% file: basicformalism.tex
Consider a set of $N$ noisy sample points  $\mathds{X} = \{\boldsymbol{x}_1,\boldsymbol{x}_2,\ldots,\boldsymbol{x}_\mathcal{D}\}$ in $\mathcal{D}$-dimensional Cartesian coordinates that denote an imaged grain boundary. A subset $\mathds{X}_S \subset \mathds{X}$ of these points is tagged as a `perimeter' that samples the edges of the grain boundary feature, with the same grid resolution as the interior. For example, in three dimensions, these points could represent a two-dimensional boundary including all edge points, or a one-dimensional triple line with its terminating quad points. We also specify a connectivity for every point in $\mathds{X}$, described by a graph Laplacian matrix $L^{(0)}$:
\begin{equation}
	L^{(0)}_{ij} = \begin{cases}
	\hfill	N(i)		\hfill	&	\text{If $i = j$}	\\
	\hfill	-I(j; i)	\hfill	&	\text{if $i \neq j$}
	\end{cases}
	\label{eq.Laplacian}
\end{equation}
where $N(i)$ is the number of points connected to $\bs{x}_i$ and $I(j;i)$ is an indicator function that is $1$ if point $j$ is connected to point $i$ and $0$ otherwise. We require that all $\boldsymbol{x}_i \in \mathds{X}_S$ remain constrained to their initial positions while the $\boldsymbol{x}_i \in \mathds{X} - \mathds{X}_S$ are smoothed, all the while adhering to the same node connectivity. We denote this smoothing operation notionally by $SMOOTH\left(\mathds{X}, \mathds{X}_S\right)$.
%, \mathds{X}_S\right)$, which modifies the position of all $\boldsymbol{x}_i \in \mathds{X} - \mathds{X}_S$, leaving all $\boldsymbol{x}_i \in \mathds{X}_S$ unchanged. 
%We detail the exact means of achieving this constrained smoothing in practice momentarily.

~~As a general rule, we enter points $\boldsymbol{x}_i$ into our hierarchy such that all $\boldsymbol{x}_i \in \mathds{X}_S$ are at one level above all $\boldsymbol{x}_i \in \mathds{X} - \mathds{X}_S$. Notationally the hierarchy level or `rank' is denoted by a function $H(\boldsymbol{x}_i)$ such that $H(\mathds{X}_S) = 1 + H(\mathds{X} - \mathds{X}_S)$; the sole purpose of $H$ being to distinguish points of different ranks and the actual returned value being a matter of choice. 
\begin{figure}[H]
	\centering
	\includegraphics[width=0.6\textwidth]{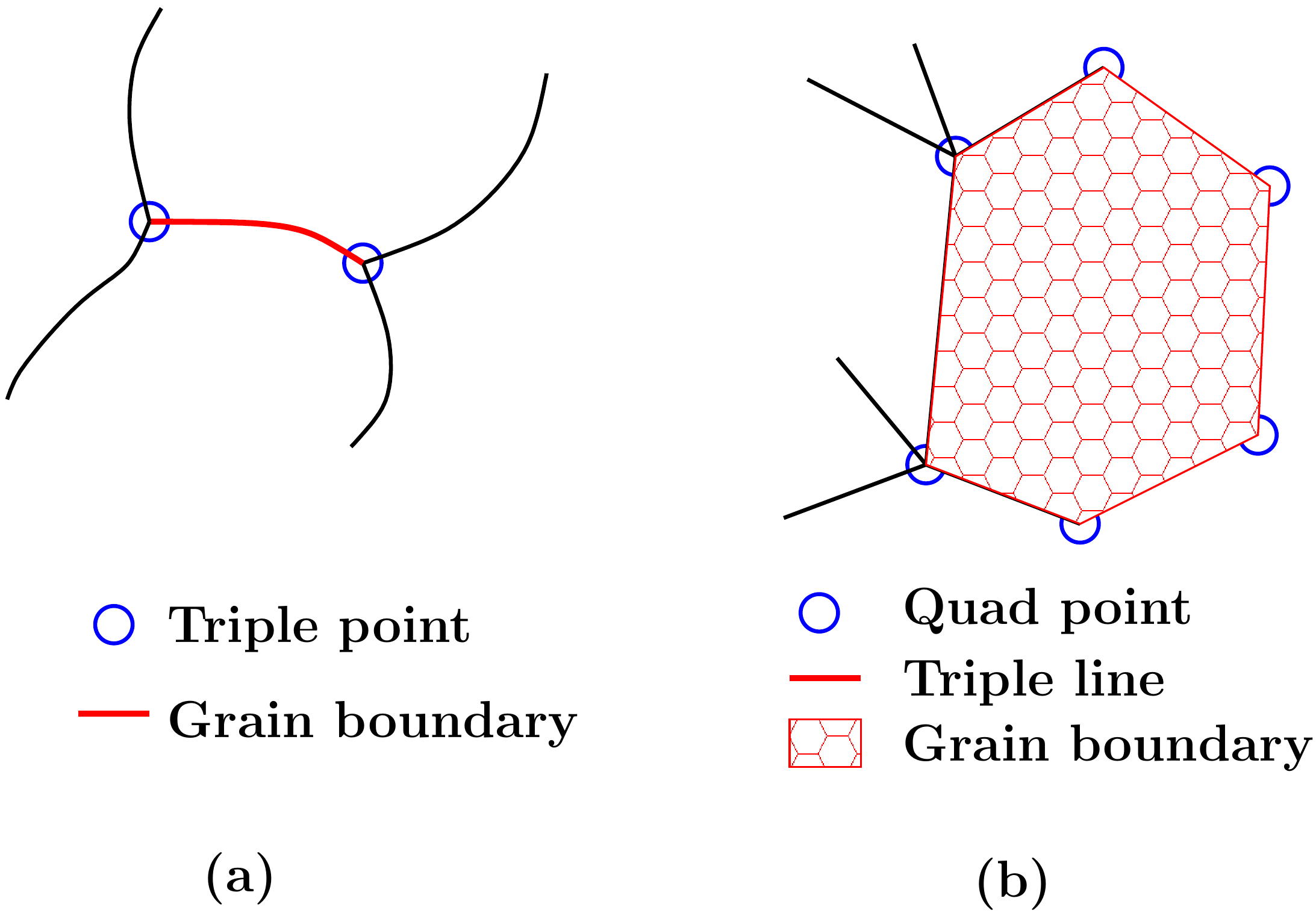}
	\caption{\textbf{(a)} Hierarchy for a two-dimensional microstructure in which all sample points in the interior of the grain boundary belong to $\mathds{X} - \mathds{X}_S$ and the triple points belong to $\mathds{X}_S$; \textbf{(b)} Three-dimensional microstructure in which the interior points of the grain boundary belong to $\mathds{X}-\mathds{X}_S$ while the boundary perimeter points belong to $\mathds{X}_S$. The perimeter points themselves can be seen to belong to a union of two-dimensional hierarchies of the type described in \textbf{(a)}.}
	\label{fig.TopHierarchy}
\end{figure}
~~Figure~\ref{fig.TopHierarchy} visualizes two common systems with different hierarchy sizes. 
Keeping in mind that in an interface network in $\mathcal{D}$-dimensional space there exist in general objects of dimensionality $d = 0, 1, \ldots, \mathcal{D}-1$, we define the rank function $H(\boldsymbol{x}_i) \equiv \mathcal{D} - d$, where $d$ corresponds to the lowest-dimensional object in the network to which $\boldsymbol{x}_i$ belongs. For example, a triple point in a 2-dimensional image is assigned a rank of $2$ because it is a zero-dimensional object, while a grain boundary interior point has a rank of $1$.  
\begin{table}[H]
	\caption{Hierarchy table for a $2$-dimensional network}
	\center
	\begin{tabular}{|c|c|c|}
		\hline
		Type of $\boldsymbol{x}_i$	&	$d$	&	$H(\boldsymbol{x}_i)$	\\	\hline
		Triple point			& 	0 	& 	2					\\	
		boundary interior		& 	1	&	1					\\	\hline
	\end{tabular}
	\label{tab.Hierarchy2D}
\end{table}
\begin{table}[H]
	\caption{Hierarchy table for a $3$-dimensional network}
	\center
	\begin{tabular}{|c|c|c|}
		\hline
		Type of $\boldsymbol{x}_i$	&	$d$	&	$H(\boldsymbol{x}_i)$	\\	\hline
		Quad point				& 	0 	& 	3					\\	
		Triple line				&	1	&	2					\\
		Boundary interior		& 	2	&	1					\\	\hline
	\end{tabular}
	\label{tab.Hierarchy3D}
\end{table}
Tables~\ref{tab.Hierarchy2D} and~\ref{tab.Hierarchy3D} show the general rule that for a given network, $d_i + H(\boldsymbol{x}_i) = \mathcal{D}$. We note that the feature of a topological element that decides its rank is its dimensionality rather than its name. For example, if a quad line existed in a network for which $\mathcal{D} = 3$ (\emph{i.e.} intersection of four grain boundary surfaces) its rank would be $2$. Based upon these definitions, the smoothing algorithm for a set of $N$ interface points in a $\mathcal{D}$-dimensional network is as follows:
\input{smoothingalgorithm.tex}
~~In summary, points of rank $N_h$ are smoothed while holding in place all previously smoothed connected points of rank $N_h' > N_h$, with highest-rank points essentially undergoing unconstrained smoothing (since $FIX$ is initially an empty set). If the highest rank elements have $d = 0$ as do quad points when $\mathcal{D}=3$ or triple points when $\mathcal{D}=2$, then one can skip `smoothing' them altogether. This scheme gives the aforementioned topological features their due importance relative to one another. The prerequisite of having sample points labeled according to topological feature is readily achievable by nearest neighbor-based clustering algorithms~\cite{Groeber2014}, which is in fact the very information represented in the \texttt{NodeType} dataset in a typical DREAM.3D microstructure file.

%% file: smoothingalgorithm.tex
\begin{algorithmic}
	\State \textbf{START}
	\State $N_h \gets \emph{Max. rank in hierarchy}$
	\State $MOV \gets \{\},~FIX \gets \{\}$
	\While{$N_h > 0$}
		\State $MOV \gets MOV \cup \{\boldsymbol{x}_i \left| H(\boldsymbol{x}_i) = N_h\right.\}$
		\State $SMOOTH\left(MOV,FIX\right)$
		\State $FIX \gets FIX \cup MOV$
		\State $N_h \gets N_h - 1$
	\EndWhile
	\State \textbf{return} $MOV$
	\State \textbf{STOP}
\end{algorithmic}

%% file: constrainedsmoothing.tex
$SMOOTH(\mathds{X}, \mathds{X}_S)$ is based on a nonparametric regression that involves penalizing, in Cartesian component-by-component fashion, the displacement between each estimated smoothed point and its \emph{unsmoothed} neighbors. If $M$ of $N$ initial points are mobile ($M < N$), a measure of the nearest neighbor fluctuations of each Cartesian component $s_i$ of $\boldsymbol{x}_i \in \mathds{X}$ can be estimated with $\left| L\boldsymbol{\sigma} + \mathbf{s}^{(b)}\right|^2$, where $\boldsymbol{\sigma} \equiv \left[s_1~s_2~\ldots~s_M\right]^T$ represents a vector of only the $s_i$ that require smoothing, $L$ is a modified graph Laplacian operator expressing the connectivity of the mobile nodes and $\mathbf{s}^{(b)}$ denotes constants that are determined from the remaining $\boldsymbol{x}_i \in \mathds{X}_S$. $\mathbf{s}^{(b)}$ in fact specifies the Dirichlet boundary conditions to Laplace's equation. Specific examples of $L$ and $\mathbf{s}^{(b)}$ are described presently. In the case of no constraints, $M = N$, $\mathds{X}_S$ is an empty set and $L$ is the full graph Laplacian introduced in Equation~\eqref{eq.Laplacian}. 
$SMOOTH$ performs simultaneous filtering of each component $s_i \rightarrow \chi_i$ by negotiating a tradeoff between fidelity to the raw data and minimization of fluctuations between \emph{smoothed} neighbors through a scalar control parameter $\epsilon$. A control function $F(\boldsymbol{\chi})$ is defined to this end:
\begin{align}
	F( \boldsymbol{\chi}) &= (1 - \epsilon) \left| \boldsymbol{\chi} - \boldsymbol{\sigma}\right|^2 + \epsilon \left|L\boldsymbol{\chi} + \mathbf{s}^{(b)}\right|^2	\label{eq.ControlFunction}  \\
	\text{where~}0 &\leq \epsilon \leq 1	\notag
\end{align}
Here $\boldsymbol{\chi} \equiv \left[\chi_1~\chi_2~\ldots~\chi_M\right]^T$ represents the array corresponding to $\boldsymbol{\sigma}$ that is further along in the smoothing process. At the extreme $\epsilon$-values of $0$ and $1$, the minimizer $\boldsymbol{\chi}_{opt}(\epsilon)$ of $F(\boldsymbol{\chi})$ respectively favors complete data fidelity ($\boldsymbol{\chi} = \boldsymbol{\sigma}$) and complete Laplace-smoothing ($L^TL\boldsymbol{\chi} + L^T\mathbf{s}^{(b)} = 0$). We further define an objective function that penalizes fluctuations between each smoothed point and its nearest \emph{unsmoothed} neighbors based on the connectivity specified in the full $N \times N$ graph Laplacian $L^{(0)}$:
\begin{equation}
	F_{obj}(\boldsymbol{\chi}(\epsilon)) = \sum_{i = 1}^N \left|\sum_{\{j \left| L^{(0)}_{ij}=-1\right.\}} \chi_i - \sigma_j\right|^2
	\label{eq.ObjectiveFunction}
\end{equation}
Crucially, we require that the minimizer of $F_{obj}$ be reached by always satisfying the optimality condition of the control function in~\eqref{eq.ControlFunction} with respect to $\boldsymbol{\chi}$ and therefore indirectly through variation of the parameter $\epsilon$ alone. This makes the smoothing operation on the $s_i$ a one-dimensional minimization in $\epsilon$ that can easily be achieved by a binary search in the interval $\left[0, 1\right]$. Briefly, the objective function $F_{obj}$ is the actual quantity being minimized in the regression, but the path taken in the objective function landscape is decided by the control function $F$. 

We define the matrix $\mathbf{x}_0$ of unsmoothed starting points as having $N$ rows and $\mathcal{D}$ columns where $\mathcal{D}$ is the dimensionality of the points. Similarly we define the identically-sized matrix $\boldsymbol{\chi}^{(0)}$ as the solution resulting from applying $SMOOTH$ to $\mathbf{x}_0$. We define $D$ and $A$ as the diagonal and adjacency matrices respectively of $L^{(0)}$. We rely on the following intermediate definitions to obtain the reduced Laplacian and constant matrices:
\begin{enumerate}
	\item	Let the integer set $\mathds{I}$ denote the indices of the points that remain fixed (\emph{i.e.} $\mathds{I} \equiv \{i \left| \boldsymbol{x}_i \in \mathds{X}_S\right.\}$ or equivalently $\mathds{X}_S \equiv \{\boldsymbol{x}_i \left| i \in \mathds{I}\right.\}$) 
	\item	If for an integer $N > 0$, $\mathds{S} = \{n_1,~n_2,\ldots\}$ is an integer set such that $ 1 \leq n_i \leq N~\forall n_i \in \mathds{S}$, then let $\widetilde{\mathds{S}} \equiv \{ i \in \mathds{Z} \left| 1 \leq i \leq N,~ i \notin \mathds{S}\right.\}$, \emph{i.e.} the complement of $\mathds{S}$ with respect to $N$.
	%If $\mathds{S} = \{i \in \mathds{Z} \left| 1 \leq i \leq N\right.\}$ \emph{i.e.} the non-negative integer set with elements less than or equal to $N$, then let $\widetilde{\mathds{S}}$ be the complement of $\mathds{S}$ \emph{i.e.} $\widetilde{\mathds{S}} = \{i \in \mathds{Z} \left| 1 \leq i \leq N,~i \notin \mathds{S}\right.\}$. 
	\item	Let the submatrix of a matrix $\mathds{M}$ formed by:
	\begin{itemize}
		\item the rows whose indices are in $\mathds{S}$ be denoted by $SM_{\text{rows}}\left(\mathds{M}, \mathds{S}\right)$.
		\item the rows \emph{and} columns whose indices are in $\mathds{S}$ be denoted by $SM_{\text{both}}\left(\mathds{M}, \mathds{S}\right)$.
	\end{itemize}
\end{enumerate}
	then the reduced Laplacian and constant matrix are defined:
\begin{align}
	L &= SM_{\text{both}}\left(L^{(0)}, \widetilde{\mathds{I}}\right)
	\label{eq.ReducedLaplacian} \\
	\mathbf{s}^{(b)} &= SM_{\text{rows}}\left(R\mathbf{x}_0, \widetilde{\mathds{I}}\right)	\label{eq.Constant}			\\
	\text{where }R\text{ is defined by: } R_{ij} &=
	\begin{cases}
		\hfill	L^{(0)}_{ij}	\hfill	& 	\text{if $j \in \mathds{I}$}	\\
		\hfill	0				\hfill	&	\text{otherwise}
	\end{cases}	\notag
\end{align}
%where $\circ$ denotes the Hadamard product. $L$ is formed from the columns of $L^{(0)}$ whose indices are not in $\mathds{I}$ while $\mathbf{s}^{(b)}$ is formed from the remaining columns. Similarly $\mathbf{x}$ is formed from the rows of $\mathbf{x}_0$ that are not in $\mathds{I}$. 
For example, if $N = 5$ points $\{\boldsymbol{x}_i = \left[x_i~y_i~z_i\right]^T \left| x_i,y_i,z_i \in \mathds{R},~i = 1, 2,\ldots,5\right.\}$ in $\mathcal{D} = 3$ dimensions are to be smoothed in which the $\boldsymbol{x}_i$ are connected sequentially with $\boldsymbol{x}_1$ and $\boldsymbol{x}_5$ to be fixed, then:
\begin{align}
	\mathds{I} &= \{1,~5\},~\widetilde{\mathds{I}} = \{2,~3,~4\}	\notag	\\
	\mathbf{x}_0 &= \left[ \boldsymbol{x}_1~\boldsymbol{x}_2~\boldsymbol{x}_3~\boldsymbol{x}_4~\boldsymbol{x}_5\right]^T_{3 \times 5},~ 
	\mathbf{x} = SM_{\text{rows}}\left(\mathbf{x}_0, \widetilde{\mathds{I}}\right) = \left[ \boldsymbol{x}_2~\boldsymbol{x}_3~\boldsymbol{x}_4\right]^T_{3 \times 3}	\notag 	\\
	L^{(0)} &= \left[\begin{array}{ccccc}
		1	&	-1	&	0	&	0	&	0	\\
		-1	&	2	&	-1	&	0	&	0	\\
		0	&	-1	&	2	&	-1	&	0	\\
		0	&	0	&	-1	&	2	&	-1	\\
		0	&	0	&	0	&	-1	&	1
	\end{array}\right],~
	L = \left[\begin{array}{ccc}
		2	&	-1	&	0	\\
		-1	&	2	&	-1	\\
		0	&	-1	&	2	
	\end{array}\right]		\notag	\\
	\mathbf{s}^{(b)} &= \left[
	\begin{array}{ccc}
		-x_1	&	-y_1	&	-z_1	\\
		0		&	0		&	0		\\
		-x_5	&	-y_5	&	-z_5	
	\end{array}\right]	\notag	\\
	D_{ij} &= \begin{cases}
		\hfill	L^{(0)}_{ij}	\hfill	& \text{If $i = j$}	\\
		\hfill	0	\hfill	& \text{Otherwise}
	\end{cases}		\notag	\\
	A_{ij} &= \begin{cases}
		\hfill	L^{(0)}_{ij}	\hfill	& \text{if $j = i \pm 1$}	\\
		\hfill	0				\hfill	& \text{Otherwise}
	\end{cases}		\notag	\\
	\boldsymbol{\chi}^{(0)} &\equiv \left[\boldsymbol{x}_1~\boldsymbol{\chi}^T~\boldsymbol{x}_5\right]^T	\label{eq.FinalComplete}
\end{align}
Equation~\eqref{eq.FinalComplete} denotes the full smoothed solution including the constrained points, with $\bs{\chi}$ as defined earlier. If $\boldsymbol{\chi}^{(0)}$ and $\boldsymbol{\chi}$ respectively satisfy the free-boundary and constrained Laplace equations, then is it clear that $L^{(0)}\boldsymbol{\chi}^{(0)} = 0_{5\times 3}$ and $L \boldsymbol{\chi} + \mathbf{s}^{(b)} = 0_{3\times 3}$. The smoothing problem is stated more compactly as the following optimization problem:
\begin{align}
	F_{obj}(\boldsymbol{\chi}^{(0)}) &= \Tr \left[ \left(D\boldsymbol{\chi}^{(0)} - A\mathbf{x}_0\right)^T \left(D\boldsymbol{\chi}^{(0)} - A\mathbf{x}_0\right) \right]
	\label{eq.ObjFunCompact}	\\
%	\boldsymbol{\chi}(\epsilon) &= \arg\min_{\mathbf{y}} \left[(1-\epsilon)\left(\mathbf{y}-\mathbf{x}\right)^T\left(\mathbf{y}-\mathbf{x}\right) + \epsilon\left(L\mathbf{y} + \mathbf{s}^{(b)}\right)^T\left(L\mathbf{y} + \mathbf{s}^{(b)}\right)\right]	
	\boldsymbol{\chi}(\epsilon) &= \left[(1-\epsilon) \mathds{1} + \epsilon L^TL\right]^{-1} \left( (1-\epsilon) \mathbf{x} - \epsilon L^T \mathbf{s}^{(b)}\right)	\label{eq.ConstraintCompact}	\\
	\epsilon_{opt} &= \arg\min_{\epsilon} F_{obj}\left(\boldsymbol{\chi}^{(0)}(\epsilon)\right)		\notag \\
	\boldsymbol{\chi}^{(0)}_{opt} &= \boldsymbol{\chi}^{(0)}(\epsilon_{opt})						\notag
\end{align}

We note from Equation~\eqref{eq.ConstraintCompact}, which is derived from the minimizer of the control function in Equation~\eqref{eq.ControlFunction}, that $\boldsymbol{\chi}(\epsilon)$ is an $M \times \mathcal{D}$ matrix and that in Equation~\eqref{eq.ObjFunCompact} the argument of the trace operator is a $\mathcal{D} \times \mathcal{D}$ symmetric matrix with non-negative eigenvalues (the case of zero eigenvalues implies that the sample points are flattened in at least one dimension). Significantly, the trace of this matrix and therefore the objective function itself represents the aggregate squared Euclidean distance of each node from its unsmoothed neighbors. The equivalence of Equations~\eqref{eq.ObjectiveFunction} and~\eqref{eq.ObjFunCompact} is seen in the simple one-dimensional smoothing example ($\mathcal{D} = 1$): $\boldsymbol{\sigma}^{(0)} = \left[\ldots\sigma_{N-1}~\sigma_N~\sigma_{N+1}\ldots\right]^T \longrightarrow \boldsymbol{\chi}^{(0)} = \left[\ldots\chi_{N-1}~\chi_N~\chi_{N+1}\ldots\right]^T$. Each column in in the $N \times \mathcal{D}$-matrix $D\boldsymbol{\chi}^{(0)} - A\mathbf{x}_0$ corresponds to one such Cartesian component in the sample frame of reference and each element is of the form $2\chi_N - \left(\sigma_{N-1} + \sigma_{N+1}\right)$. This is precisely the argument of the $\left|\cdot\right|^2$ operation in \eqref{eq.ObjectiveFunction}. The objective function in \eqref{eq.ObjFunCompact} represents the operation in \eqref{eq.ObjectiveFunction} being performed simultaneously on all $\mathcal{D}$ Cartesian components. Minimizing them simultaneously is completely equivalent to minimizing the (reference frame-invariant) trace of the matrix $\left(D\boldsymbol{\chi}^{(0)} - A\mathbf{x}_0\right)^T \left(D\boldsymbol{\chi}^{(0)} - A\mathbf{x}_0\right)$.

The $SMOOTH$ algorithm is finally given by:
\input{smoothing.tex}
The resulting surface consisting of the smoothed points with the preserved original connectivity is nonparametric. Qualitatively, the algorithm attempts to determine the least jagged surface passing in between the sample points, thus maintaining data fidelity. This precludes a major problem in applying iterative Laplace-like techniques, that of over- or under-smoothing. For the applications of $SMOOTH$ in the remainder of this text, the threshold value of $\partial F_{\text{obj}} / \partial \epsilon$  was taken to be $10^{-7}$.

We point out that the smoothing scheme outlined in this section allows users to define the components of the algorithm for specific requirements. For instance, users of mesh smoothing algorithms like finite element method (FEM) might want to explicitly incorporate mesh quality metrics into the objective function as an alternative to remeshing. This in turn may well decide the manner of stepping in the objective function landscape and therefore shape the control function. Our smoothing paradigm permits the flexibility of user-defined objective and control functions, all the while heeding the hierarchy between the components of the grain boundary network. We address the applicability of smoothed meshes obtained from the objective function in Equation~\eqref{eq.ObjFunCompact} to finite element applications in Section~\ref{S:meshquality}.

%% file: smoothing.tex
\begin{algorithmic}
	\State \textbf{START}
	\State $\epsilon\gets 0.5$, $\Delta\epsilon \gets 0.25$
	\State $\boldsymbol{\chi} \gets \left[(1-\epsilon) \mathds{1} + \epsilon L^TL\right]^{-1}\left[(1-\epsilon)\mathbf{x} - \epsilon L^T \mathbf{s}^{(b)}\right]$
	\State $\boldsymbol{\chi}^{(0)} \gets \boldsymbol{\chi} \cup \mathds{X}_S$	\Comment{\emph{i.e.} add the points that are held fixed}
	\State $F_{\text{obj}} \gets \Tr \left[\left(D\boldsymbol{\chi}^{(0)} - A\mathbf{x}_0\right)^T\left(D\boldsymbol{\chi}^{(0)} - A\mathbf{x}_0\right)\right]$
	\While{$\left|\partial F_{\text{obj}}/\partial\epsilon\right| \geq Threshold$}
		\If{$\partial F_{\text{obj}}/\partial\epsilon > 0$}
			\State $\epsilon \gets \epsilon - \Delta \epsilon$
		\Else
			\State $\epsilon \gets \epsilon + \Delta \epsilon$
		\EndIf
		\State $\Delta\epsilon \gets \Delta\epsilon/2$
		\State $\boldsymbol{\chi} \gets \left[(1-\epsilon) \mathds{1} + \epsilon L^TL\right]^{-1}\left[(1-\epsilon)\mathbf{x} - \epsilon L^T \mathbf{s}^{(b)}\right]$
		\State $\boldsymbol{\chi}^{(0)} \gets \boldsymbol{\chi} \cup \mathds{X}_S$	\Comment{\emph{i.e.} add the points that are held fixed}
%		\State $F \gets \min_\lambda \{\left|\left(D\boldsymbol{\chi}^{(0)} - A\mathbf{x}_0\right)^T\left(D\boldsymbol{\chi}^{(0)} - A\mathbf{x}_0\right) - \lambda \mathds{1}\right| = 0\}$
		\State $F_{\text{obj}} \gets \Tr \left[\left(D\boldsymbol{\chi}^{(0)} - A\mathbf{x}_0\right)^T\left(D\boldsymbol{\chi}^{(0)} - A\mathbf{x}_0\right)\right]$
	\EndWhile
	\State \textbf{return} $\boldsymbol{\chi}^{(0)}$
	\State \textbf{STOP}
\end{algorithmic}

%% file: smoothingdigitizedcurves.tex
We describe as a first demonstration of $SMOOTH$ the problem of filtering a set of pixelated points $\mathds{X} \equiv \{\left(i, y_i\right) \left|i = 1,2,\ldots,N~\text{and}~y_i \in \mathds{Z}\right.\}$ representative of a curve in a plane. We list the following general properties of such a set of points:
\begin{enumerate}
	\item	The coordinates are integral multiples of some basis of vectors in the plane. This could indicate either a square or triangular grid as implemented in commercial electron backscatter diffraction (EBSD)~\cite{Schwartz2009} software or near-field high-energy diffraction microscopy (nf-HEDM)~\cite{Lienert2011,Li2013}. 
	\item	Every point has at least one nearest neighbor in at least one direction on this integer grid. This is characteristic of discretized sampling of continuous curves and surfaces in general.
\end{enumerate}

\begin{figure}[H]
	\centering
	\includegraphics[width=\textwidth]{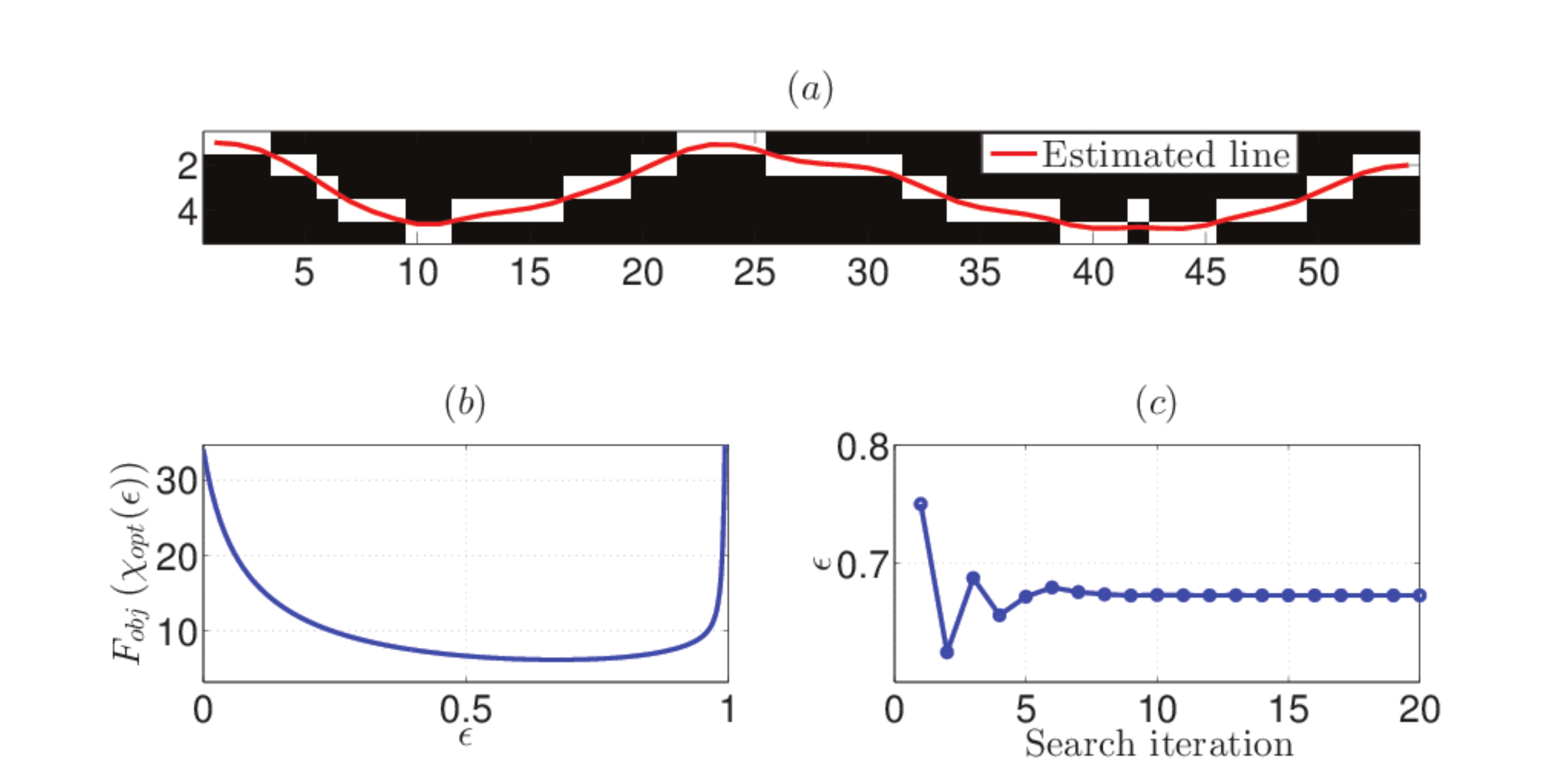}
	\caption{\textbf{(a)} Pixelated and thinned line and its smooth estimate; \textbf{(b)} A plot of the objective function in its domain $\left[0, 1\right]$ shows a shallow minimum; \textbf{(c)} Search for the optimal value of the control parameter $\epsilon$ that strikes the best balance between data fidelity and smoothness.}
	\label{fig.BinarizedLineSmooth}
\end{figure}

Such points can be obtained from pixelated images in standard formats by first generating a phase field (for instance a field of unique integers assigned to each grain), taking the magnitude of the gradient of this field and binarizing it. A morphological `skeletonizing' operation can then be applied to this binarized field~\cite{Haralick1992,Lam1992,Kong1996}. This same technology is used in the field of biometrics, for example, to thin down high-resolution images of fingerprints to features of single-pixel thickness for further analysis. For nf-HEDM images, one can collect directly the voxel (\textbf{vo}lume pi\textbf{xel}) edges that border two different grains, as decided by some segmentation criterion. In our example the coordinates of the sample points are integers on a square grid. The image of the curve has been skeletonized to ensure that each sample point has no more than two of the $8$ immediate square-grid neighbors belonging to the pixelated line (interior points have two $8$-neighbors and the terminal points have one).
The results of constrained smoothing on such a pixelated curve is shown in Figure~\ref{fig.BinarizedLineSmooth} with the perimeter points fixed at the unsmoothed grid point locations.

In Figure~\ref{fig.BinarizedLineSmooth}\textbf{(a)} the unfiltered sample points are taken to be the pixel centers. That the final estimated curve passes almost through most of the pixel corners is a consequence of the algorithm attempting to choose a smooth path between the unfiltered sample points, much like an old-fashioned wooden spline negotiating a smooth path between its fixed control points. Given a fixed boundary condition, the skeletonization of the original pixelated curve ensures that the smooth estimate stays within a pixel width of the original smooth curve. This is also seen in the smooth estimates of the two-dimensional cross-sections of grains in Figure~\ref{fig.fe2dsmoothing}\textbf{(d)}.

%% file: smoothingknownshapes.tex
\begin{figure}[H]
	\centering
	\subfloat[]{\includegraphics[height=0.225\textheight]{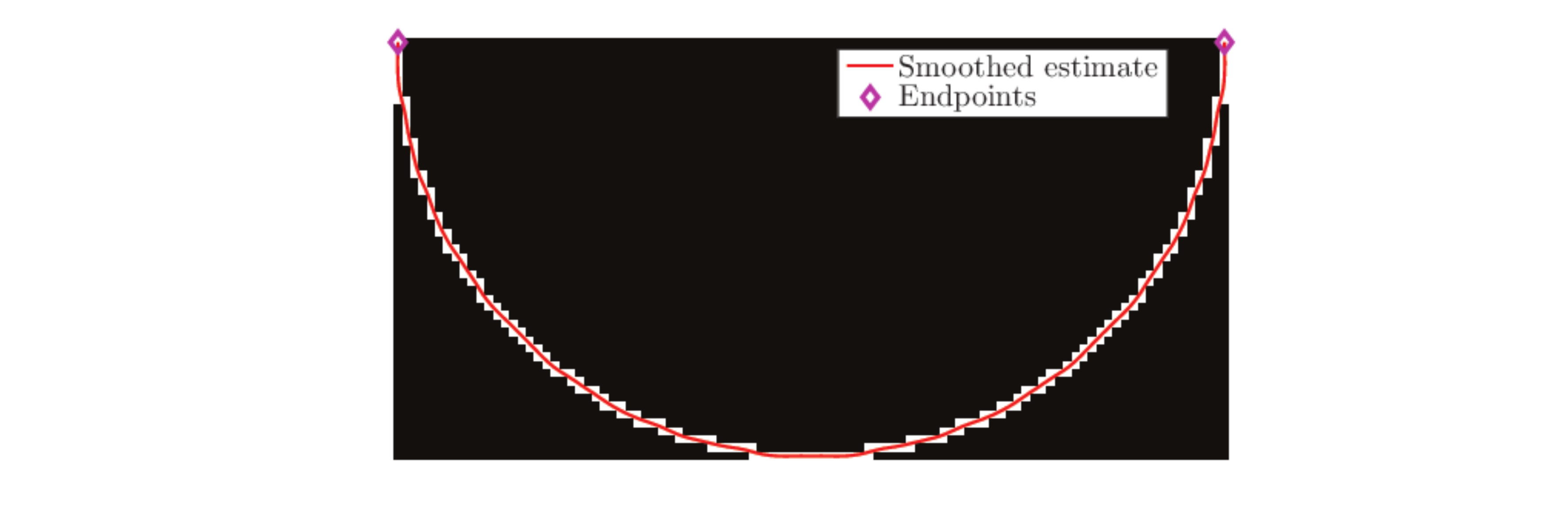}}		\vfill
	\subfloat[]{\includegraphics[height=0.225\textheight]{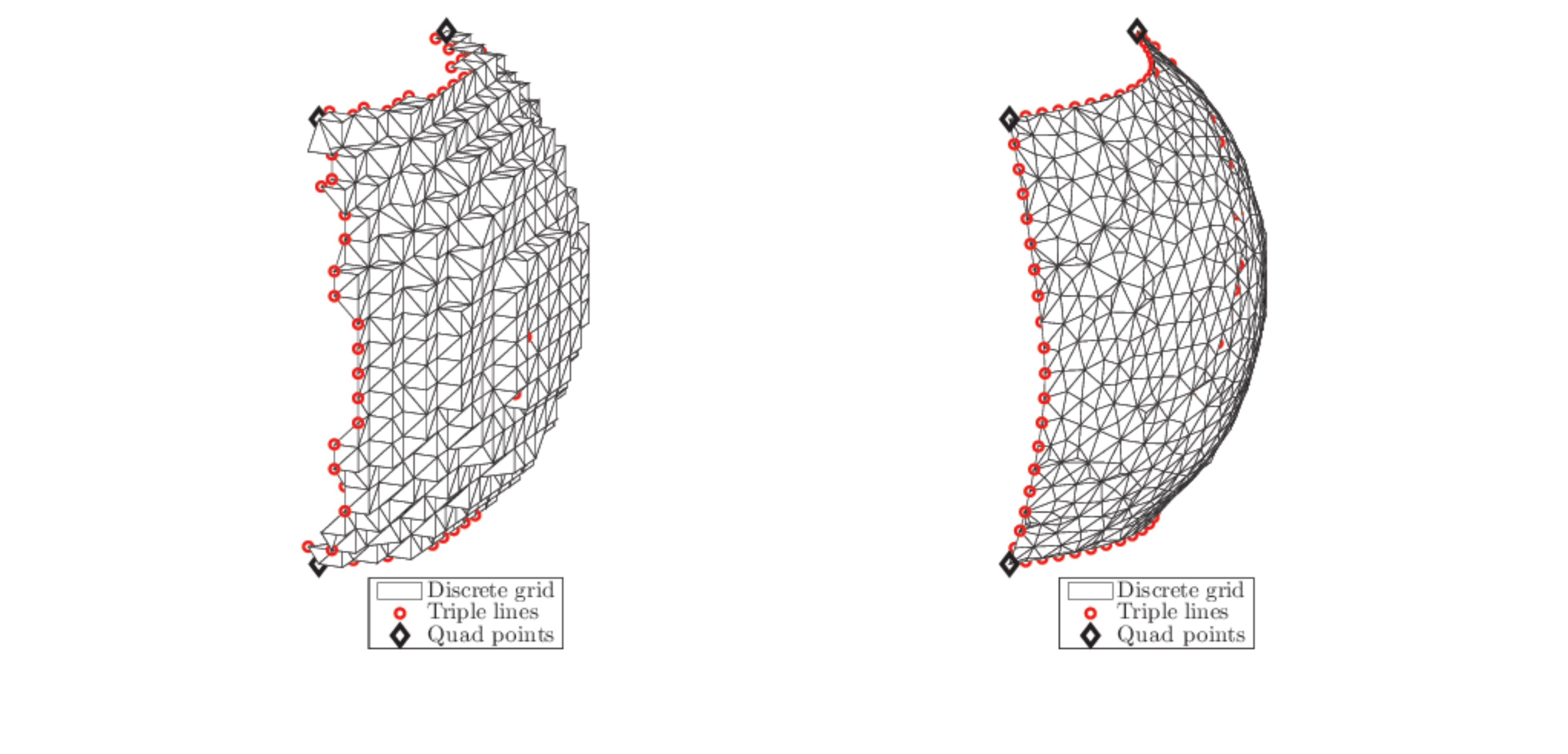}}		\vfill
	\subfloat[]{\includegraphics[height=0.225\textheight]{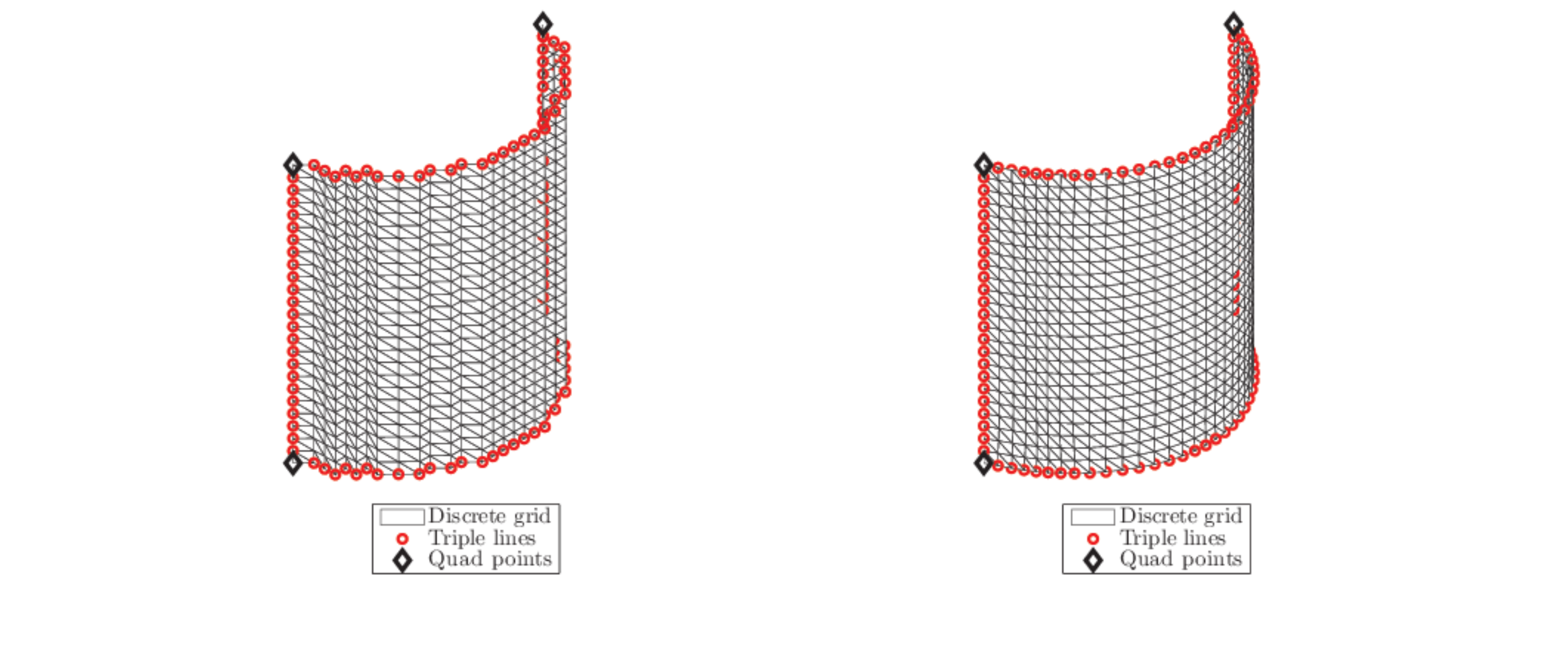}}	\vfill
	\caption{\textbf{(a)} Semicurcular arc with a $50$ pixel radius; \textbf{(b)} A discretized spherical patch ($400$ voxels per unit length) spanning equal angles ($100^\circ$) in mutually perpendicular directions and its smoothed version; \textbf{(b)} A cylindrical patch discretized to $400$ voxels per unit length and its smoothed version. }
	\label{fig.knownshapes}
\end{figure}

~~In this section the hierarchical $SMOOTH$ algorithm is applied to open surfaces that are geometry primitives in two and three dimensions. Simple parameterizations for these primitives provide a means of comparison with a smoothed solution on a point-to-point basis. We describe trends in the errors for different primitives as a function of voxel density. Focusing attention on open surfaces allows us to simulate smoothing in the presence of topological features characteristic of a grain boundary network. The three primitives chosen are a 2D circle, a 3D sphere, and a 3D cylinder. Surface slices from these primitives were characterized by the following:
\begin{itemize}
	\item	The gridding resolution was chosen in terms of the number of voxels per unit length, $N$.
	\item	\textbf{Circle}: A semicircular arc of unit radius, whose endpoints were reset to unit radius after discretization. These endpoints were held constrained during smoothing.
	\item	\textbf{Sphere}: A square patch spanning $100^\circ$ in two mutually perpendicular directions cut out from the surface of a sphere of radius $0.03$ units, with the edges of the square treated as triple lines and the vertices as quad points. The quad points alone were constrained to lie on the sphere, while the others were subject to discretization on a cubic lattice.
	\item	\textbf{Cylinder}: A rectangular patch cut from the surface of a cylinder of radius $0.03$ units, parallel to its axis and spanning $150^\circ$ along the azimuth. The edges of the rectangle were treated as triple lines and the vertices as quad points. The quad points were constrained to remain on the surface of the cylinder.
\end{itemize}
	The choice of the sphere and cylinder radii are indicative of the typical size of a grain from earlier nf-HEDM measurements~\cite{Hefferan2010}. The quality of smoothing was expressed as the error in the estimated radius of the primitive in question. Specific examples smoothing on these primitives are shown in Figure~\ref{fig.knownshapes}. 
%\begin{figure}[H]
%	\centering
%	\includegraphics[height=0.5\textheight]{KnownShapesExamples.pdf}
%	\caption{\textbf{(a)} Semicircular arc with a $50$ voxel radius. \textbf{(b)} A discretized spherical patch ($400$ voxels per unit length) spanning equal angles in two mutually perpendicular directions along the spherical surface, and \textbf{(c)} its smoothed version. \textbf{(d)} A cylindrical patch discretized to $250$ voxels per unit length and \textbf{(e)} its smoothed version. In \emph{(b)} and \emph{(d)}, the quad points have been set to their original positions on their respective surfaces.}
%	\label{fig.knownshapes}
%\end{figure}

\begin{figure}[H]
	\centering
	\subfloat[]{\includegraphics[height=0.22\textheight]{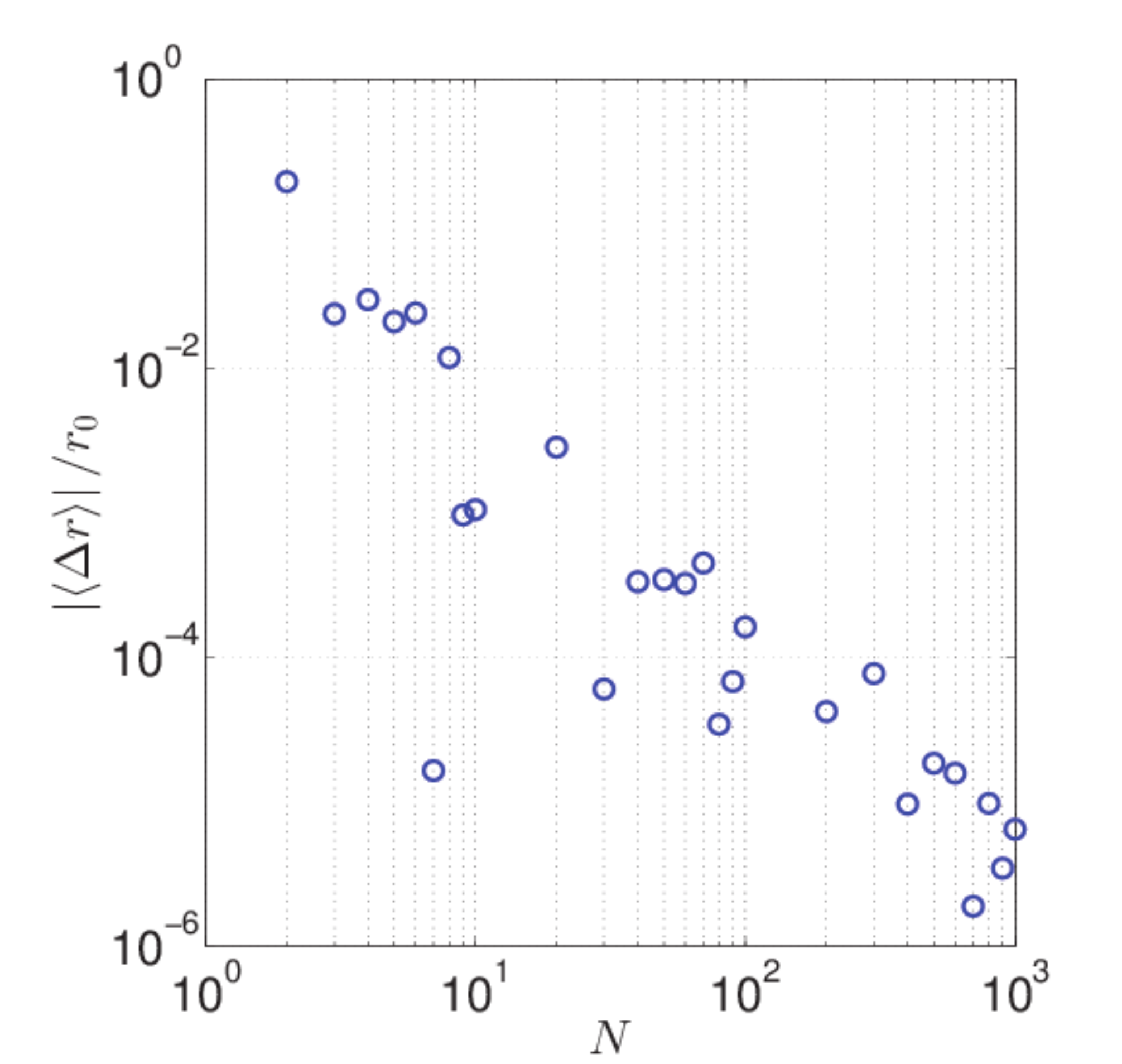}}		\hfill
	\subfloat[]{\includegraphics[height=0.22\textheight]{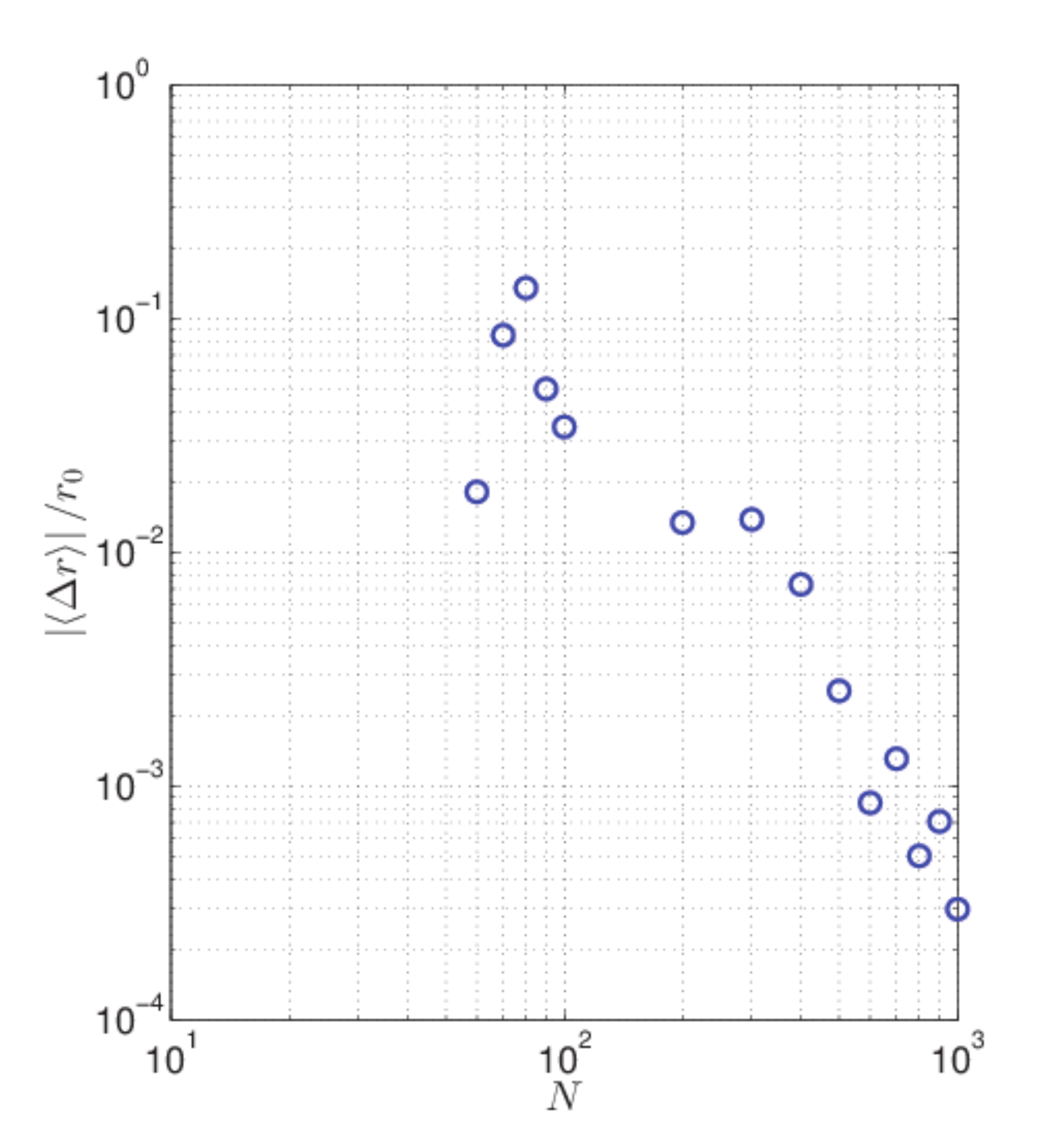}}		\hfill
	\subfloat[]{\includegraphics[height=0.22\textheight]{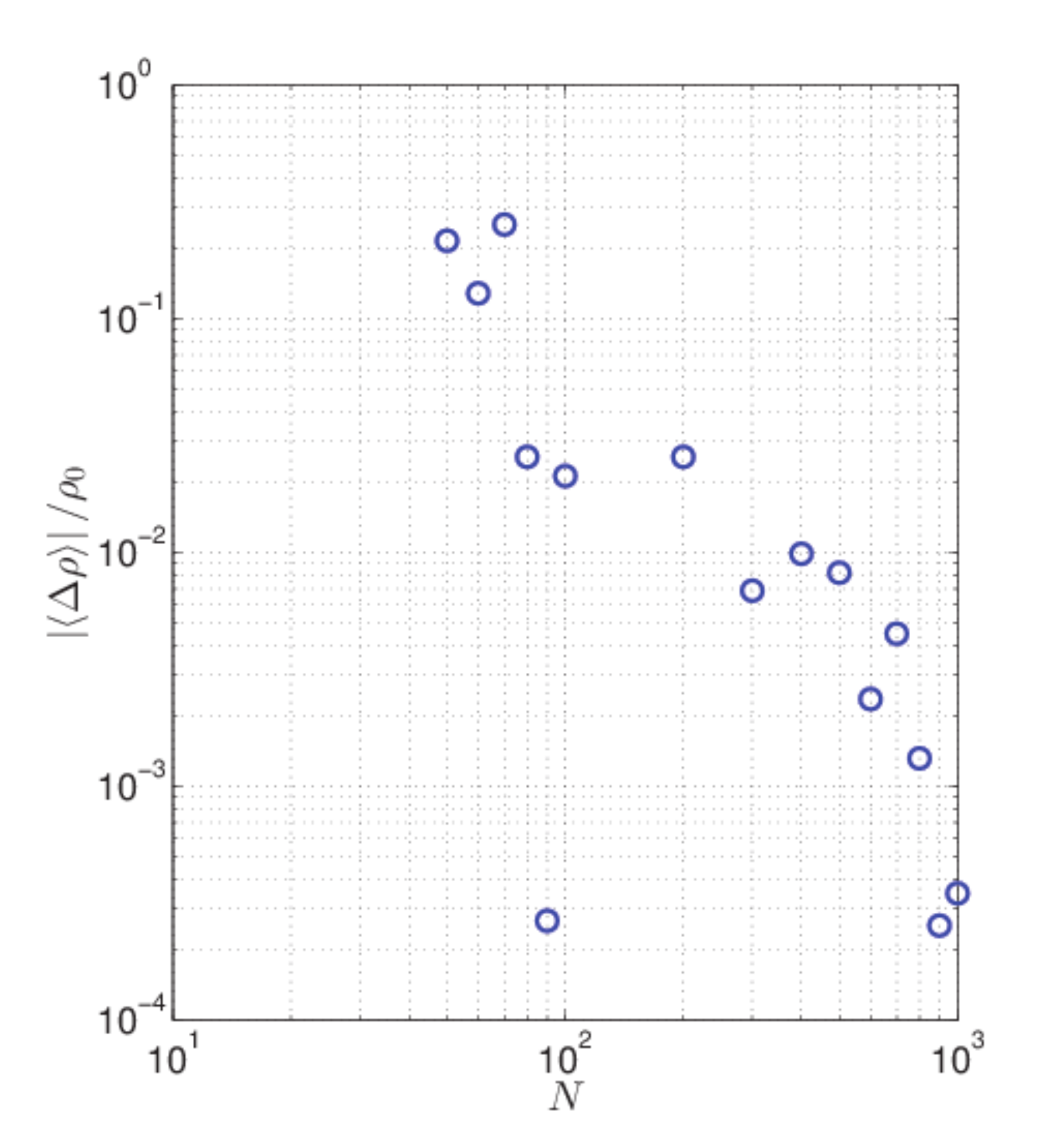}}		\hfill
	\vfill
	\subfloat[]{\includegraphics[height=0.22\textheight]{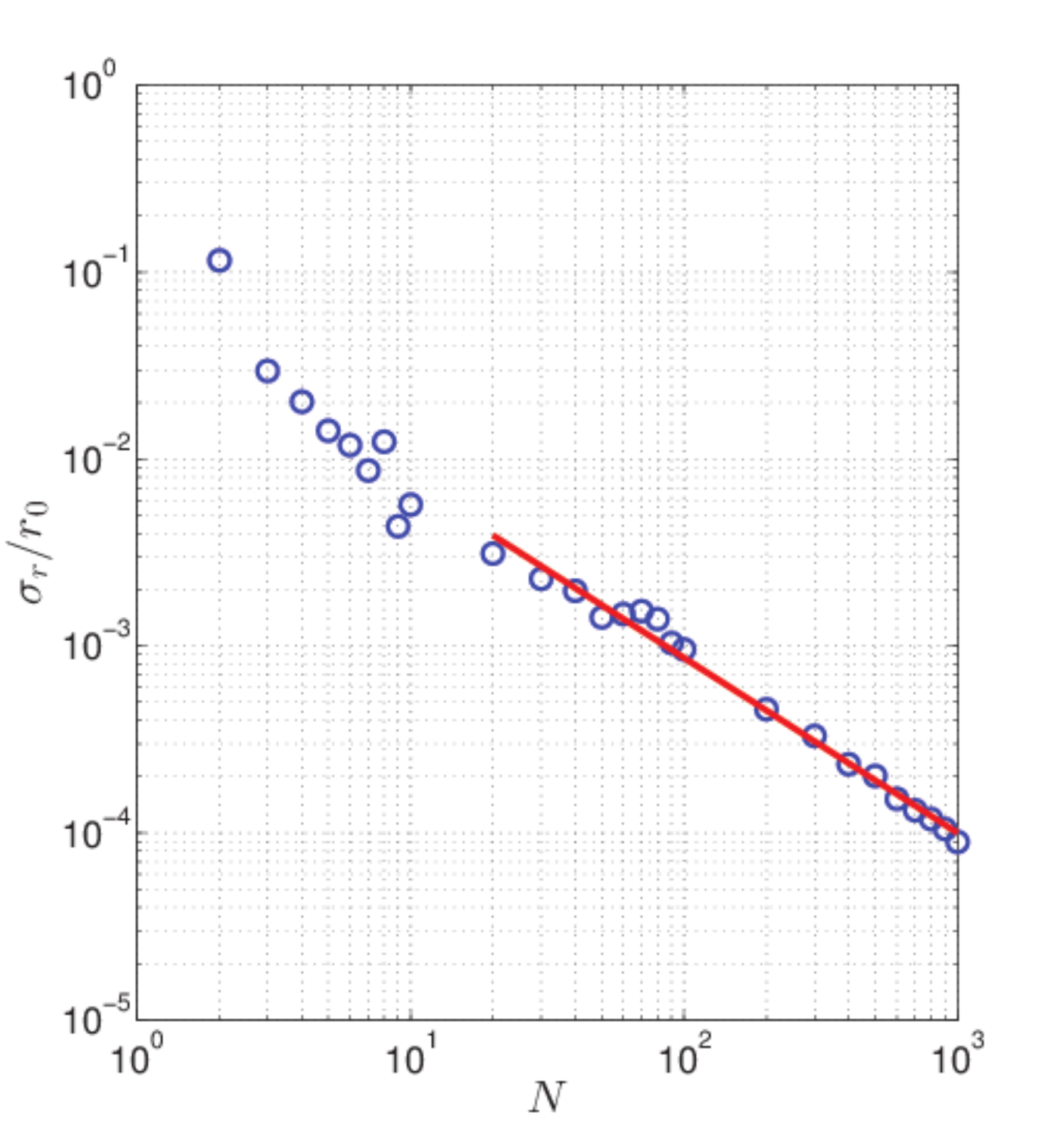}}		\hfill
	\subfloat[]{\includegraphics[height=0.22\textheight]{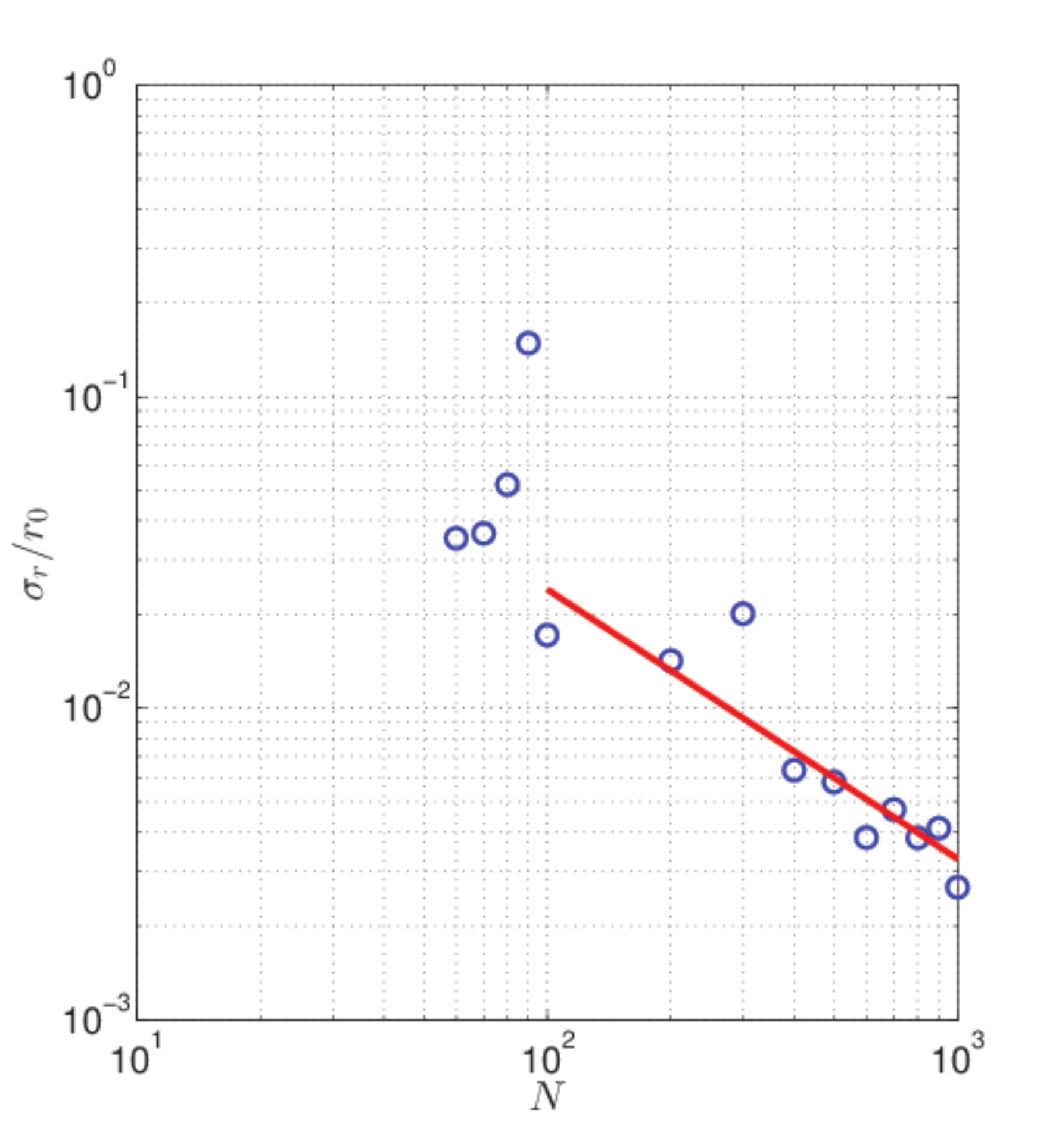}}		\hfill
	\subfloat[]{\includegraphics[height=0.22\textheight]{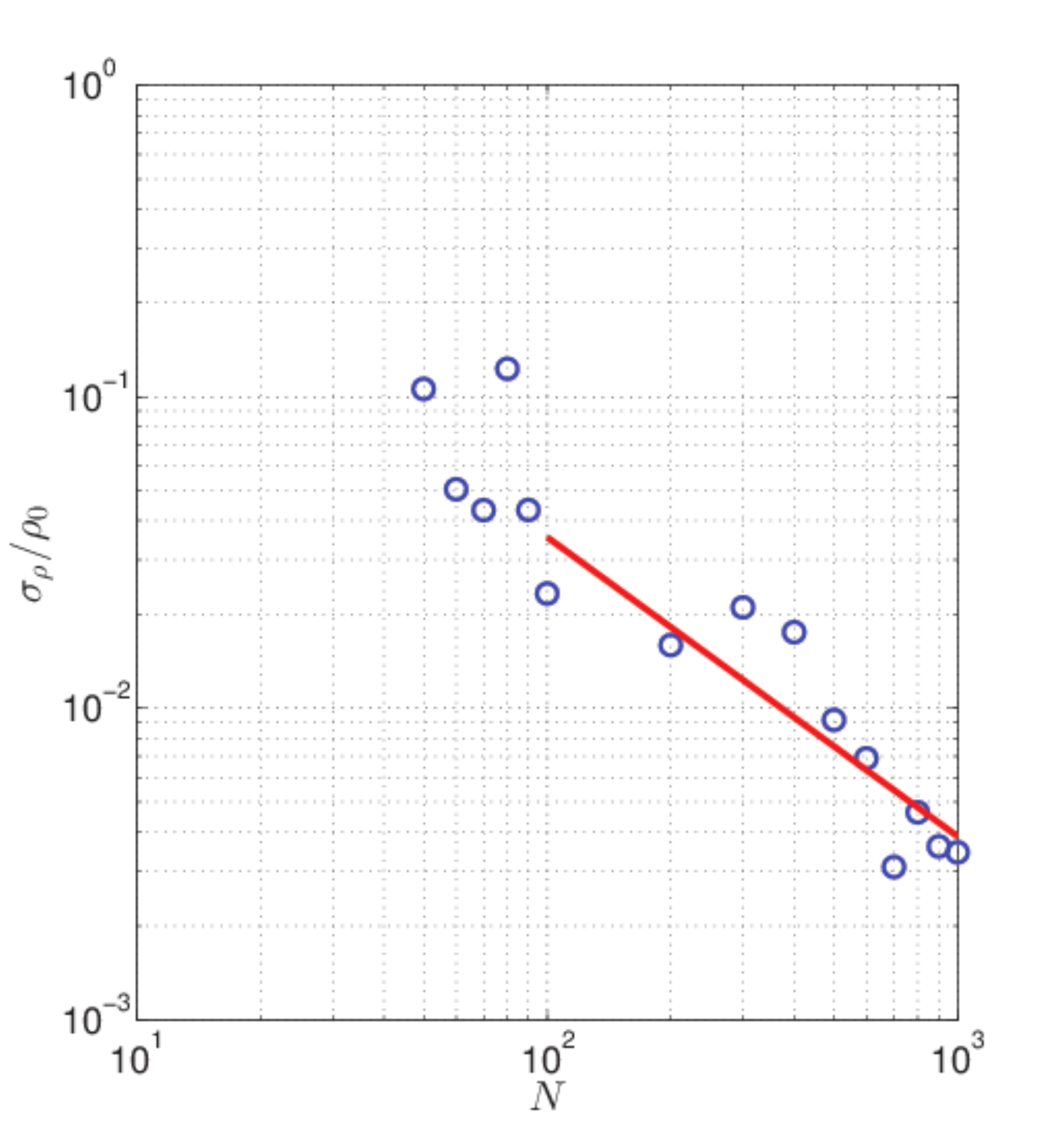}}		\hfill
	\caption{Trends in the error in estimated size $\Delta r = r - r_0$ of the \textbf{(a)} circle, \textbf{(b)} sphere and \textbf{(c)} cylinder as a function of voxel density per unit length $N$. Trends in the error spread $\sqrt{\left<\Delta r^2\right> - \left<\Delta r\right>^2}/r_0$ for the \textbf{(d)} circle, \textbf{(e)} sphere and \textbf{(f)} cylinder as a function of $N$. The red lines are fits to the power law $\sigma = 10^{p_0} N^{p_1}$ whose determined coefficients are listed in Table~\ref{tab.powerlaw}. Not all the points are used in the fits owing to coarse gridding at low values of $N$.}
	\label{fig.knownshapeserrors}
\end{figure}

~~The fidelity of the final smoothed result to the original primitive was quantified in terms of the point-to-point difference in radii of the original and smoothed surfaces. In the case of the sphere and the cylinder, the first spherical and cylindrical polar coordinates are respectively used (\emph{i.e} $r$ and $\rho$).  Shown in Figure~\ref{fig.knownshapeserrors} are trends in the estimated error $\Delta r \equiv r - r_0$ and its standard deviation $\sigma_r$, taken over the smoothed mesh nodes, for all three primitives. If the unit of length is taken to be a millimeter, the relative error in the region between $N = 300$ and $N = 1000$ is particularly relevant for techniques like nf-HEDM since they correspond to a pixel size range of $1 \mu m$ to $3.33 \mu m$, which brackets the known experimental resolution~\cite{Hefferan2010a}. For comparison, the radii of the spherical and cylindrical patches were chosen to be $30\mu m$. 
%\begin{figure}[H]
%	\centering
%	\includegraphics[width=\textwidth]{KnownShapesErrorTrends}
%	\caption{Trends in the relative deviation $\sqrt{\left<\Delta r^2\right> - \left<\Delta r\right>^2}/r_0$ for the \textbf{(a)} circle, \textbf{(b)} sphere and \textbf{(c)} cylinder as a function of the voxel density per unit length $N$. The red line is a fit to the power law $\sigma = 10^{p_0} N^{p_1}$ whose determined coefficients are listed in Table~\ref{tab.powerlaw}. Not all the points are used in the fit in \emph{(a)} because at lower values of $N$, the circle is too coarse-gridded.}
%	\label{fig.knownshapeserrors}
%\end{figure}
\begin{table}
	\caption{Estimated power law coefficients of the error trends for various primitives.}
	\center
	\begin{tabular}{|c|c|c|}
		\hline
		Primitive	&	$p_1$		&	$p_0$		\\ \hline
		Circle		&	$-0.93724$	&	$-1.18898$	\\
		Sphere		&	$-0.86535$	&	$0.11215$	\\
		Cylinder	&	$-0.96061$	&	$0.47041$	\\ \hline
	\end{tabular}
	\label{tab.powerlaw}
\end{table}
~~The power law behavior in the error trends was tested for each primitive as is seen in the straight line fits in Figure~\ref{fig.knownshapeserrors}. The coefficients $p_0$ and $p_1$ of the estimated power law  $f(N) = 10^{p_0}  \times N^{p_1}$ are listed in Table~\ref{tab.powerlaw}. Of particular interest is $p_1$ which is seen to lie close to $-1$ for all three primitives. This is simply explained by the fact that the length error is $1/N$, the size of one voxel. 
%goes as the $d^{th}$ root of the volume error, which in the case of $N$ pixels per unit length, goes as $N^{-d}$. 
The lowered fit quality for the sphere and cylinder is attributed to the difficulty in obtaining a perfect stair-stepped mesh for these primitives. We further note that the relative error for each primitive is around a fraction of a percent ($< 10^{-2}$) at the spatial resolution of nf-HEDM ($1.48 \mu m$ or $\sim 675$ pixels per millimeter).

\begin{figure}
	\centering
	\subfloat[]{\includegraphics[width=0.5\textwidth]{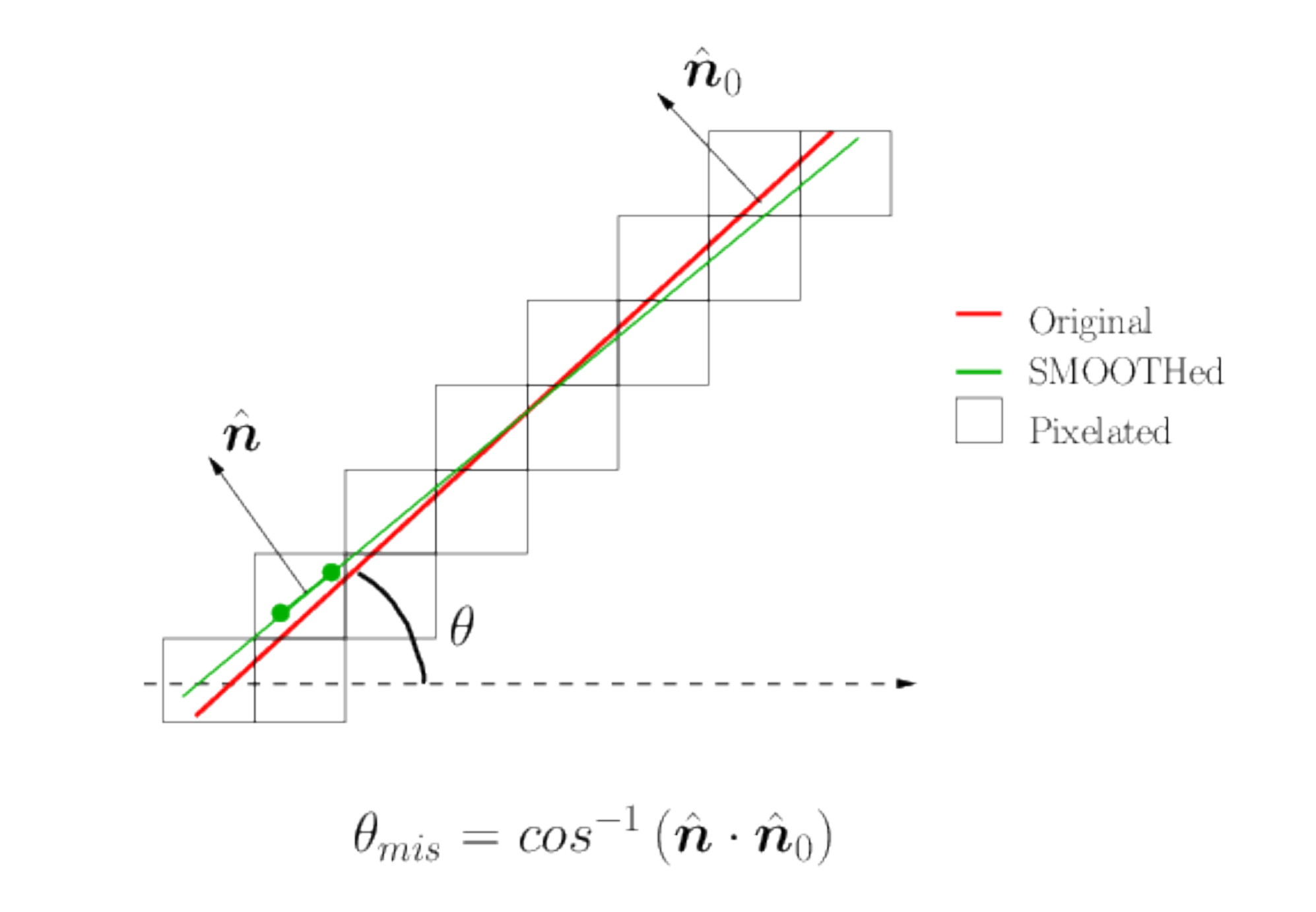}}
	\hfill
	\subfloat[]{\includegraphics[width=0.5\textwidth]{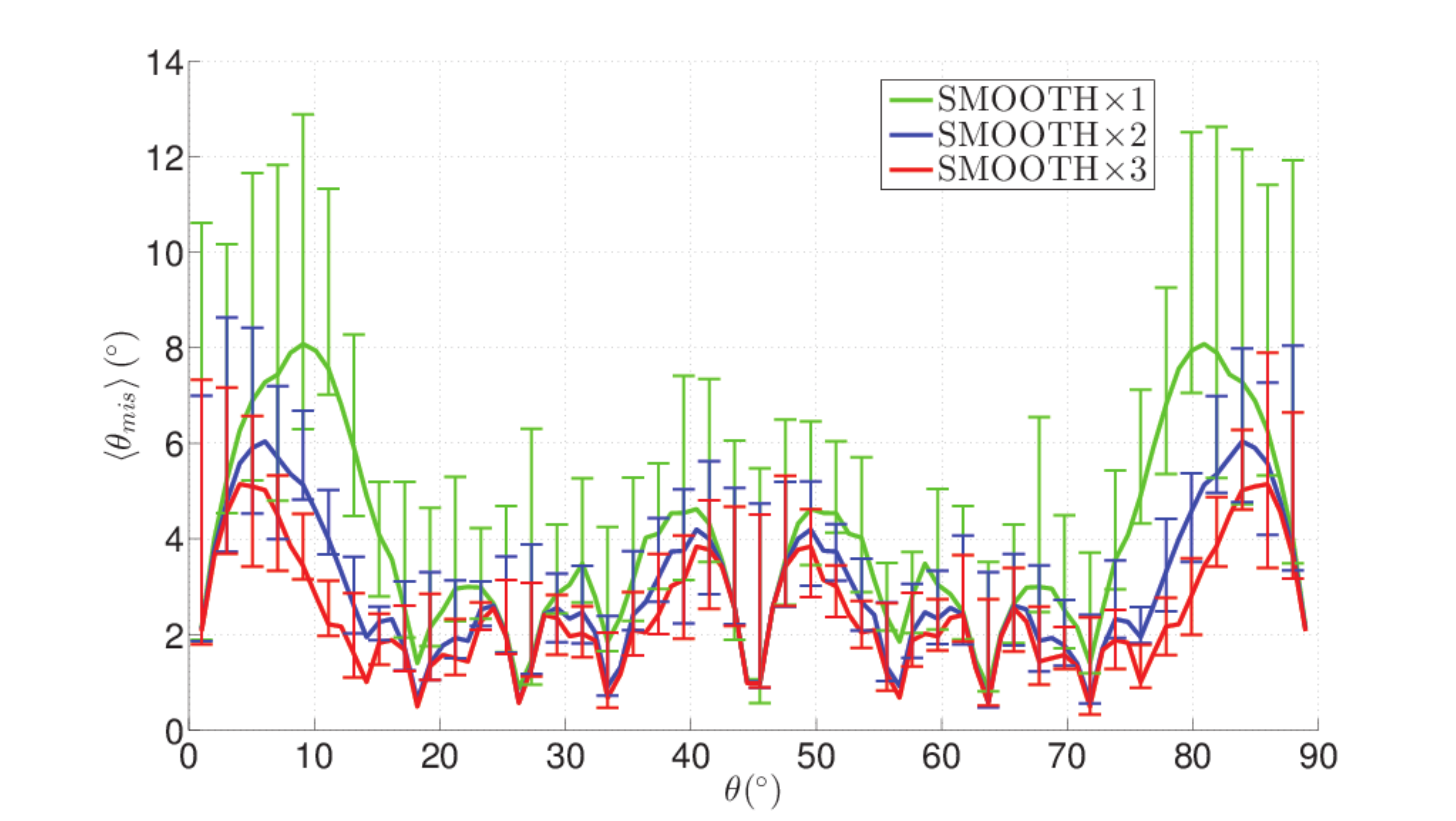}}
	\caption{Estimation of deviation in expected inclination for a discretization of $600$ pixels per unit length, for repeated applications of the $SMOOTH$ algorithm. \textbf{(a)} Schematic of a section of the simulated line, \textbf{(b)} Mean of $\theta_{mis}$ (taken over length of the entire smoothed line) as a function of inclination $\theta$. The error bars denote the standard deviations of the one-sided distribution on either side of the mean.}
	\label{fig.errors}
\end{figure}

~~Another smoothing quality metric that is easily calculated for two dimensions is the error in the local normal of a curve. This is particularly relevant to surface imaging applications. We estimate the local deviation about the known normal of the sections of a pixelated straight line that has been $SMOOTH$ed with its endpoints held at their true positions. This deviation is determined for different inclinations of the original line to the Cartesian grid. A schematic and results for the inclination range of $0^\circ$ to $90^\circ$ is shown in Figure~\ref{fig.errors}. It is seen that while the mean deviation varies with the inclination of the original line, it falls within a few degrees of the actual normal which justifies the use of $SMOOTH$ in calculations of grain boundary character distribution~\cite{Sintay2012,Khorashadizadeh2011,Ratanaphan2014}. 

%% file: twodimmicsmooth.tex
\begin{figure}
	\centering
	\includegraphics[width=0.95\textwidth]{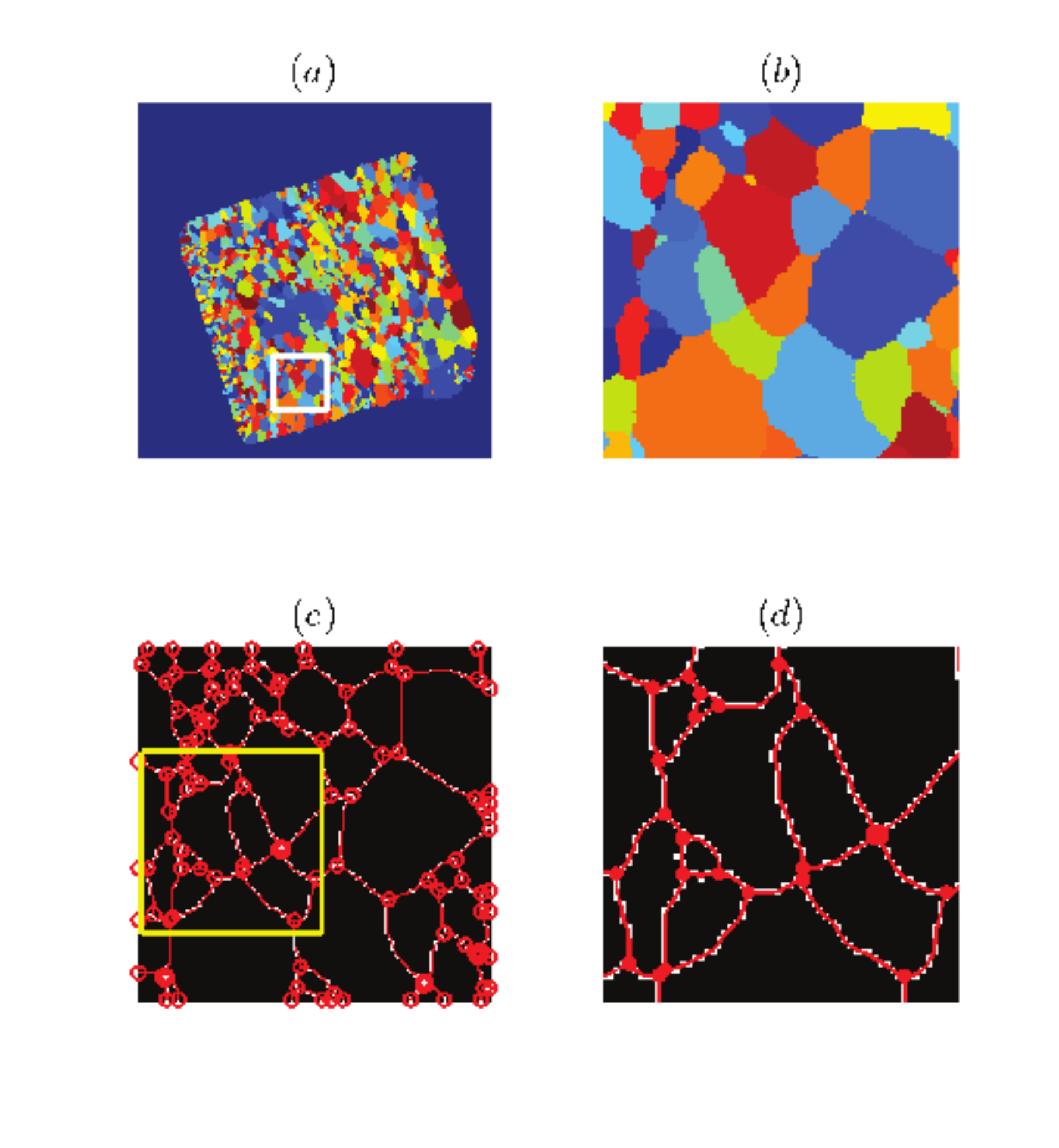}
	\caption{\textbf{(a)} Discrete phase field of a single layer of well-ordered microstructure imaged by nf-HEDM. Each grain is colored according to a unique integer assigned to it; \textbf{(b)} The region of interest in \emph{(a)} zoomed in; \textbf{(c)} Image obtained from taking the derivative of the phase field in \emph{(a)} and binarizing it above a chosen threshold, then performing a skeletonizing operation. Triple points were identified as those grain boundary points that have three distinct phase values in their $8$-neighborhood, while grain boundary interior points as those having exactly two distinct phase values in their $8$-neighborhood. On this is superposed the result of $SMOOTH$ing carried out for each boundary while holding its associated triple points fixed; \textbf{(d)} The result of the smoothing operation in the region of interest from \emph{(c)}.}
	\label{fig.fe2dsmoothing}
\end{figure}

~~The constrained smoothing is next demonstrated on a real two-dimensional microstructure imaged with nf-HEDM. 
%The material is well-ordered $\alpha$-iron with a BCC crystal structure. 
	The sample points $\mathbf{x}_i \equiv \left[x_i~y_i\right]^T$ of each grain boundary are expressed in integer units of suitable in-plane step size (in this case, $2 \mu m$) and are classified as belonging to a grain interior, grain boundary or triple point depending on the number of unique grains represented in the $8$-neighborhood . The optimization is performed simultaneously over $x_i$ and $y_i$ and therefore $\boldsymbol{\chi}^{(0)}$ is an $N \times 2$-matrix. The results on a section of microstructure are shown in Figure~\ref{fig.fe2dsmoothing}.

%% file: threedimmicsmooth.tex
\begin{figure}[H]
	\centering
	\includegraphics[width=\textwidth]{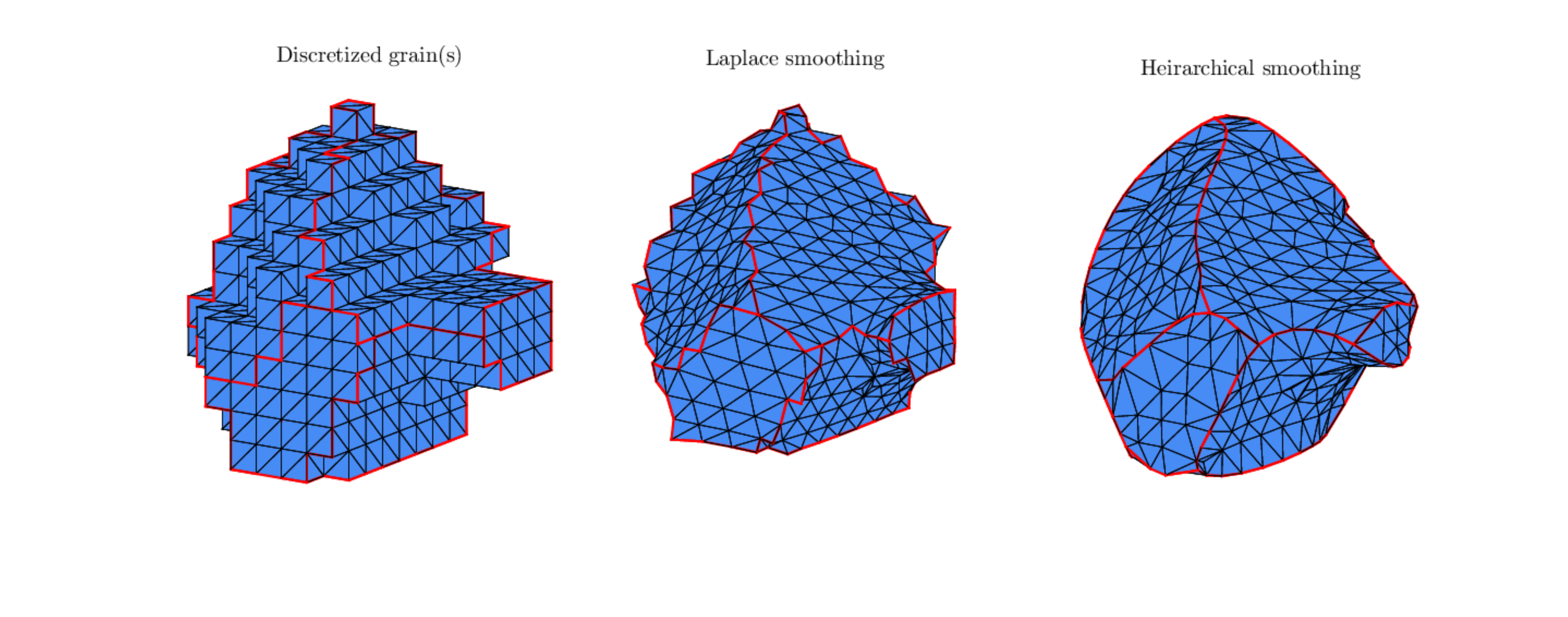}
	\caption{Discretized grain mesh smoothed using Laplace smoothing (with $400$ iterations and smoothing parameter $\lambda$ set to $0.025$, $0.5$ and $0.025$ for interior nodes, triple lines and quad points respectively), and parameter-free hierarchical smoothing. The red lines are triple lines. }
	\label{fig.3dgrain1}
\end{figure}
\begin{figure}[H]
	\centering
	\includegraphics[width=\textwidth]{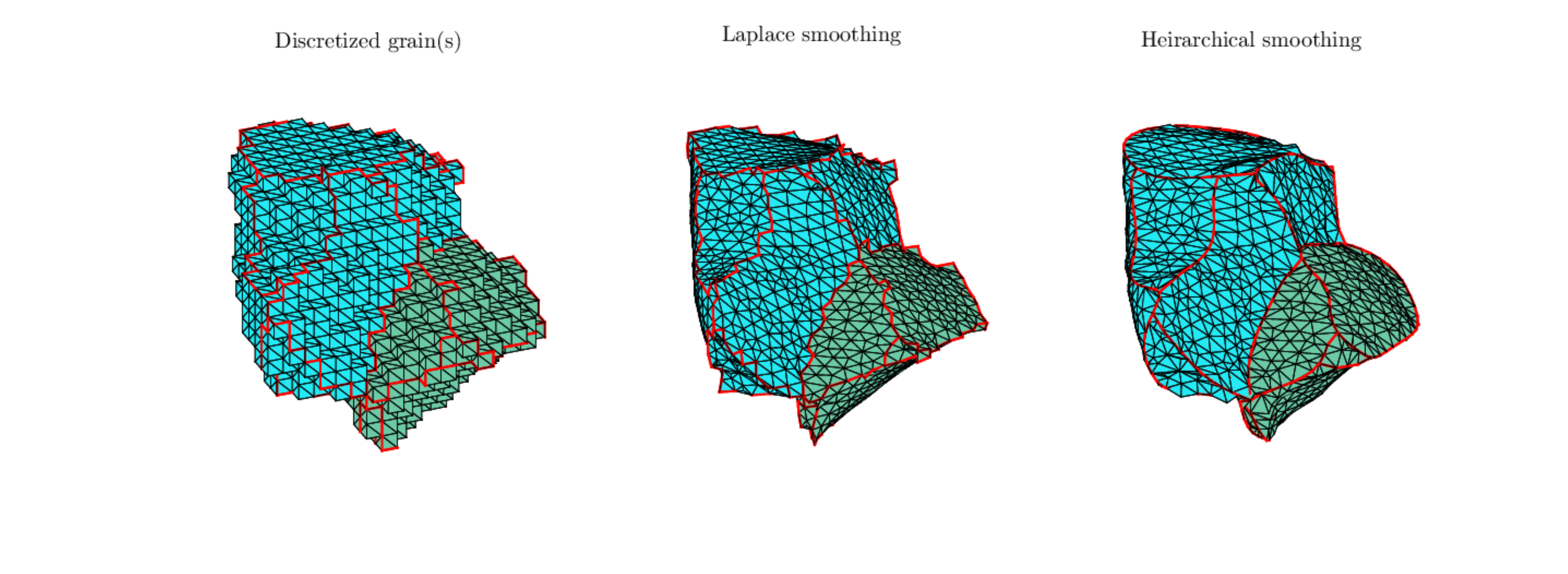}
	\caption{Discretized mesh of a pair of neighboring grains smoothed using Laplace smoothing (same parameters as in Figure~\ref{fig.3dgrain1}) and parameter-free hierarchical smoothing.}
	\label{fig.3dgrainpair}
\end{figure}
\begin{figure}[H]
	\centering
	\includegraphics[width=\textwidth]{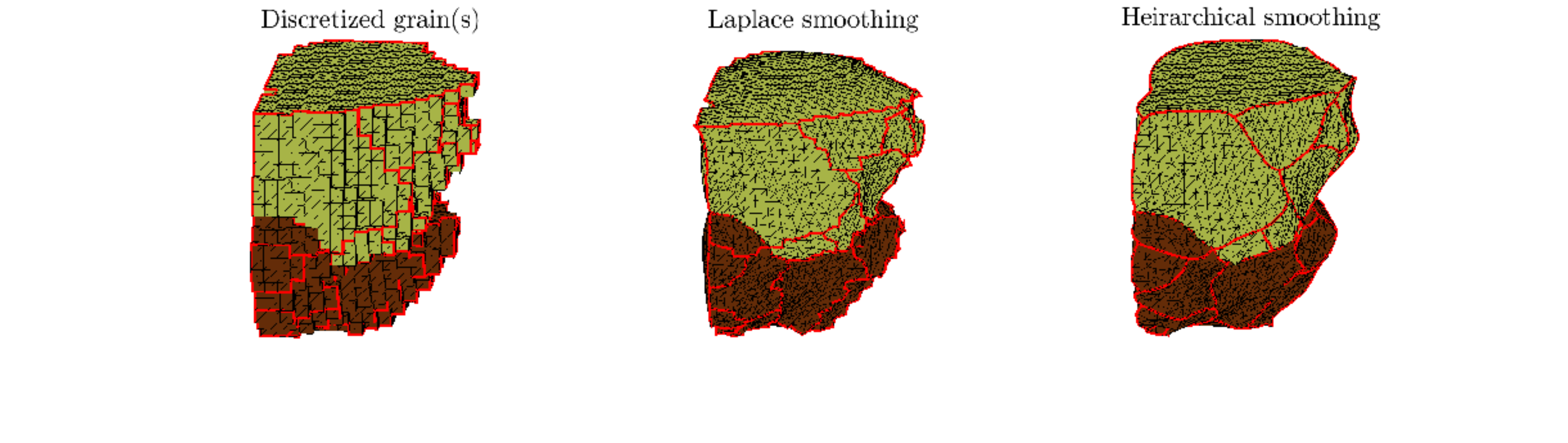}
	\caption{A more complicated grain pair structure with a larger number of topological features. In this example, the flatness of the top surface in the discretized grain owing to the grain being on the edge of the sample is preserved with hierarchical smoothing, while Laplace smoothing returns a clearly visible and unphysical bulge along that face. }
	\label{fig.3dgrainpair2}
\end{figure}

~~Hierarchical smoothing is demonstrated on select grains of a well-ordered three-dimensional microstructure measured with nf-HEDM and is compared to the results of Laplace smoothing. The prerequisite node connectivity on the grain surfaces was obtained by first segmenting the microstructure into its constituent grains and then triangulating the faces of the cubic voxels along grain boundaries~\cite{Groeber2014}. The result of this operation is a `quick-and-dirty' Delaunay mesh on stepped grain surfaces characteristic of discrete sampling. A few important points about this bookkeeping process that inform the subsequent smoothing are:
\begin{itemize}
	\item	The grain surface nodes are unambiguously classified into their topological types \emph{i.e.} boundary interiors, triple lines or quad points. It is worth mentioning that the preprocessing described above is external to the hierarchical smoothing algorithm itself and as such is not the focus of this work. 
	\item	The meshing and bookkeeping is done in such a manner as to ensure that nodes along triple lines and at quad points are shared between the neighboring topological features. 
\end{itemize}
Figures~\ref{fig.3dgrain1},~\ref{fig.3dgrainpair}~and~\ref{fig.3dgrainpair2} show comparisons of hierarchical smoothing with Laplace smoothing in which the ease of movement of the triple points is enhanced by assigning them a greater Laplace smoothing parameter $\lambda$. The values of $\lambda$ for the parent volumes for each of these grains were determined by trial and error (as a user would have to do) by visually minimizing the distortion from  the original square-gridded grain. This is a highly inefficient process that is not required in hierarchical smoothing. 

%\begin{figure}
%	\centering
%	\includegraphics[width=\textwidth]{CompareWithLaplace_Grain423}
%	\caption{Discretized grain mesh smoothed using Laplace smoothing (with $400$ iterations and smoothing paremeter $\lambda$ set to $0.025$, $0.5$ and $0.025$ for interior nodes, triple lines and wuad points respectively), and parameter-free hierarchical smoothing. The black dots on the discretized grain are the quad points and the red lines are triple lines. }
%	\label{fig.3dgrain1}
%\end{figure}

%% file: meshquality.tex
\begin{figure}
	\centering
	\subfloat[]{\includegraphics[width=.5\textwidth]{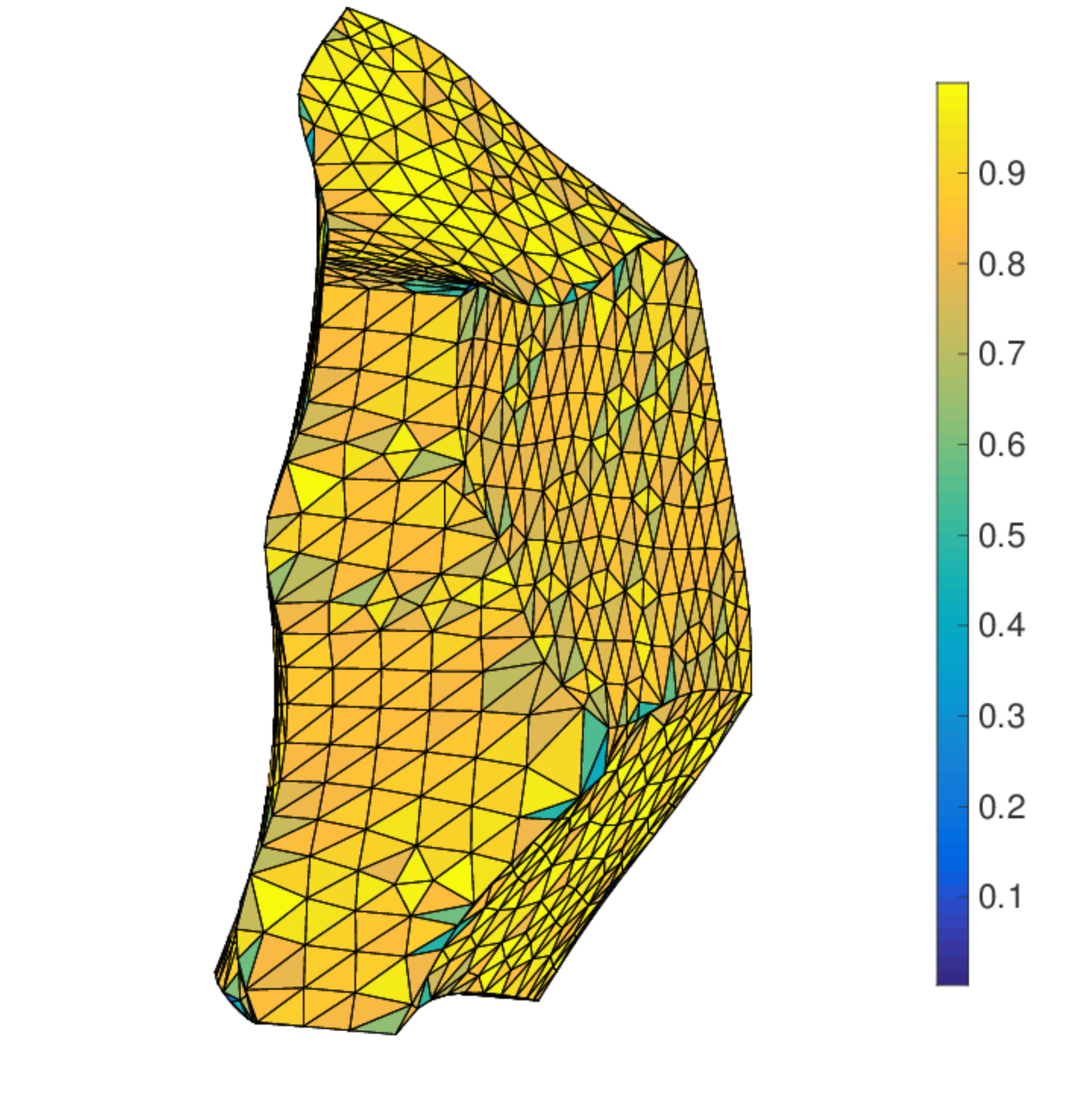}}
	\hfill
	\subfloat[]{\includegraphics[width=.5\textwidth]{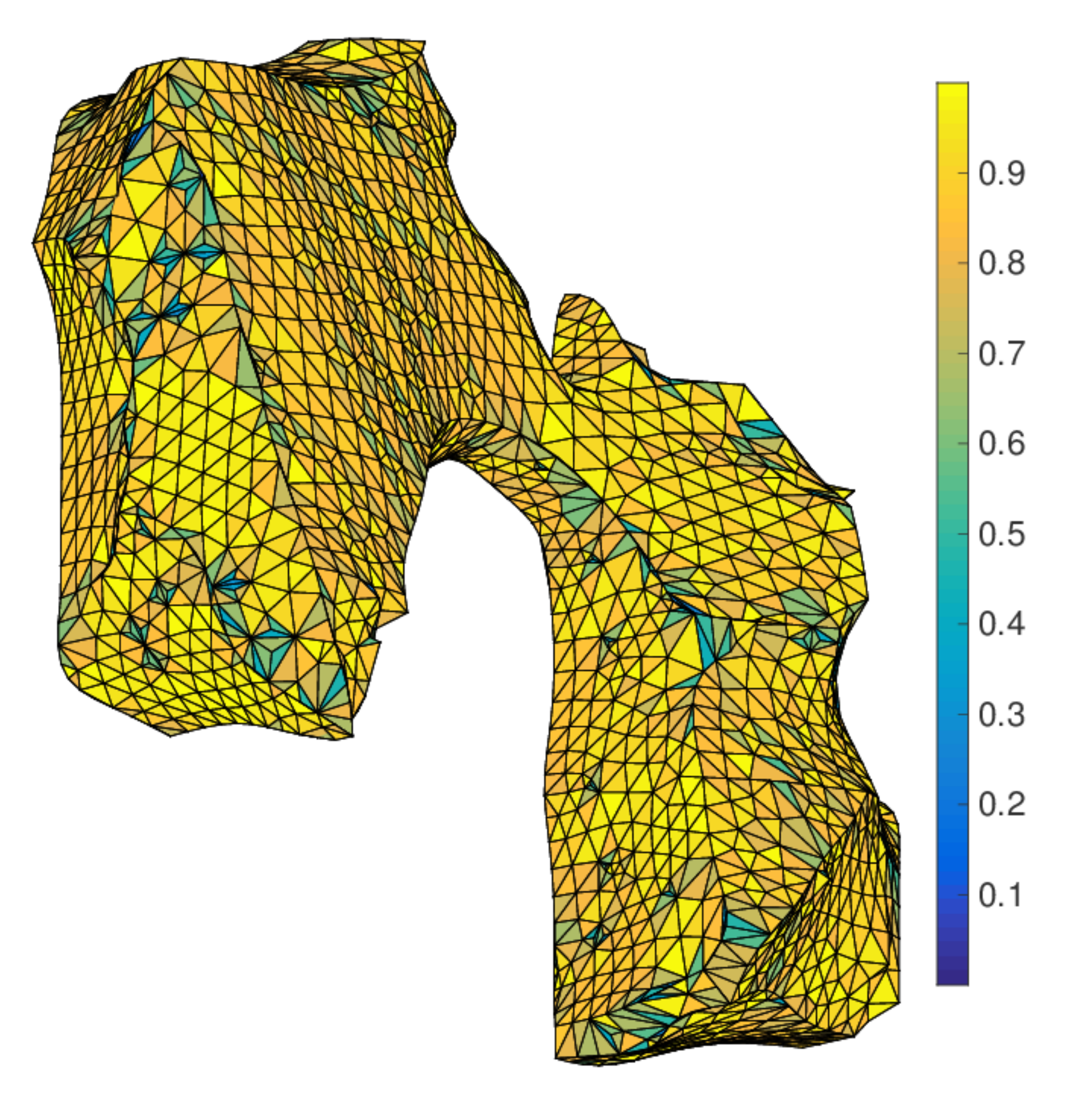}}
	\vfill
	\subfloat[]{\includegraphics[width=.5\textwidth]{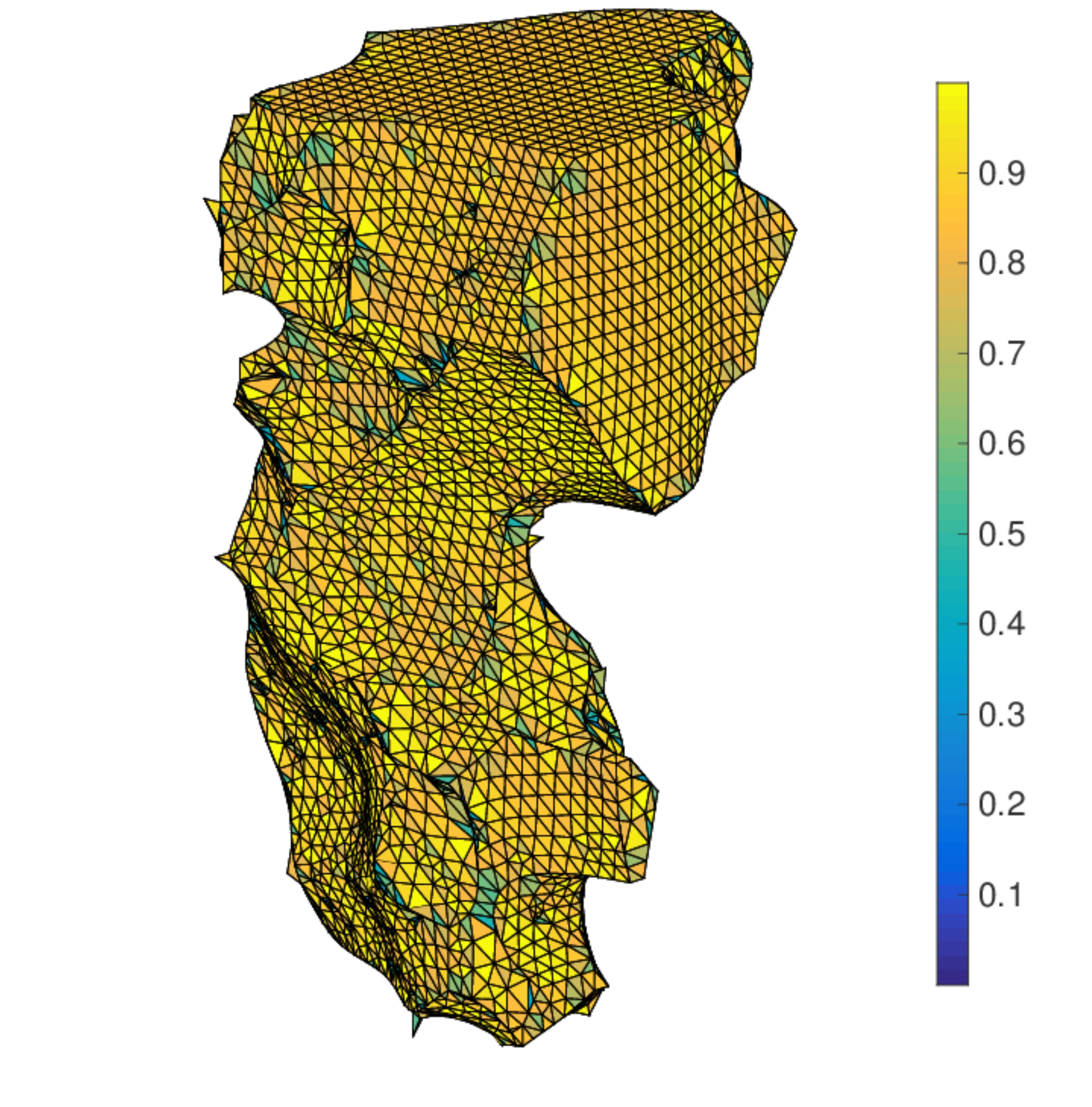}}
	\hfill
	\subfloat[]{\includegraphics[width=.5\textwidth]{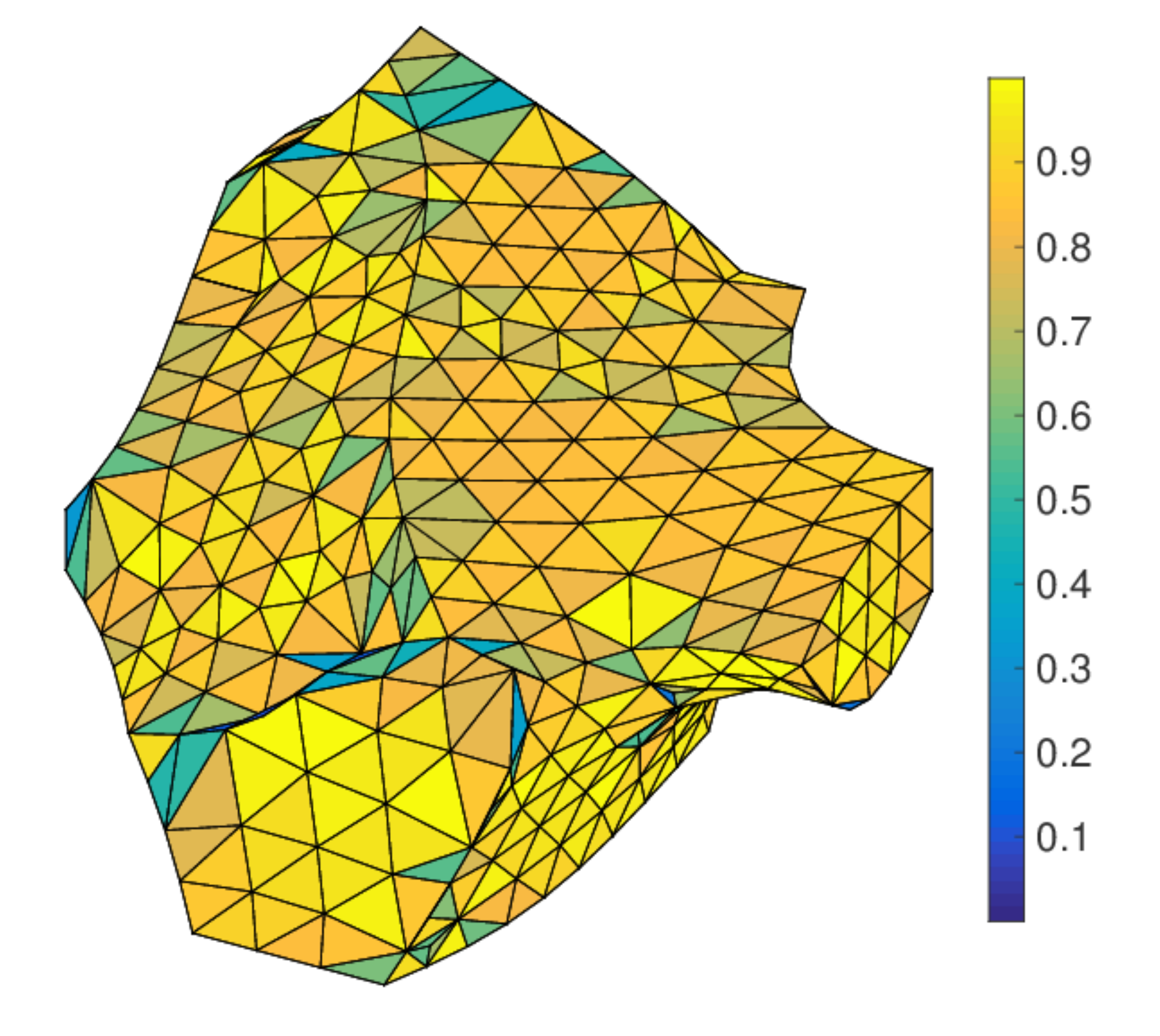}}
%	\vfill
%	\subfloat[]{\includegraphics[width=\textwidth]{VolumeMeshQuality}}
	\caption{Mesh quality of select grains in the microstructure volume. The grain in \textbf{(d)} is the same as the one in Figure~\ref{fig.3dgrain1}.}
	\label{fig.meshquality}
\end{figure}
\begin{figure}
	\centering
	\includegraphics[width=\textwidth]{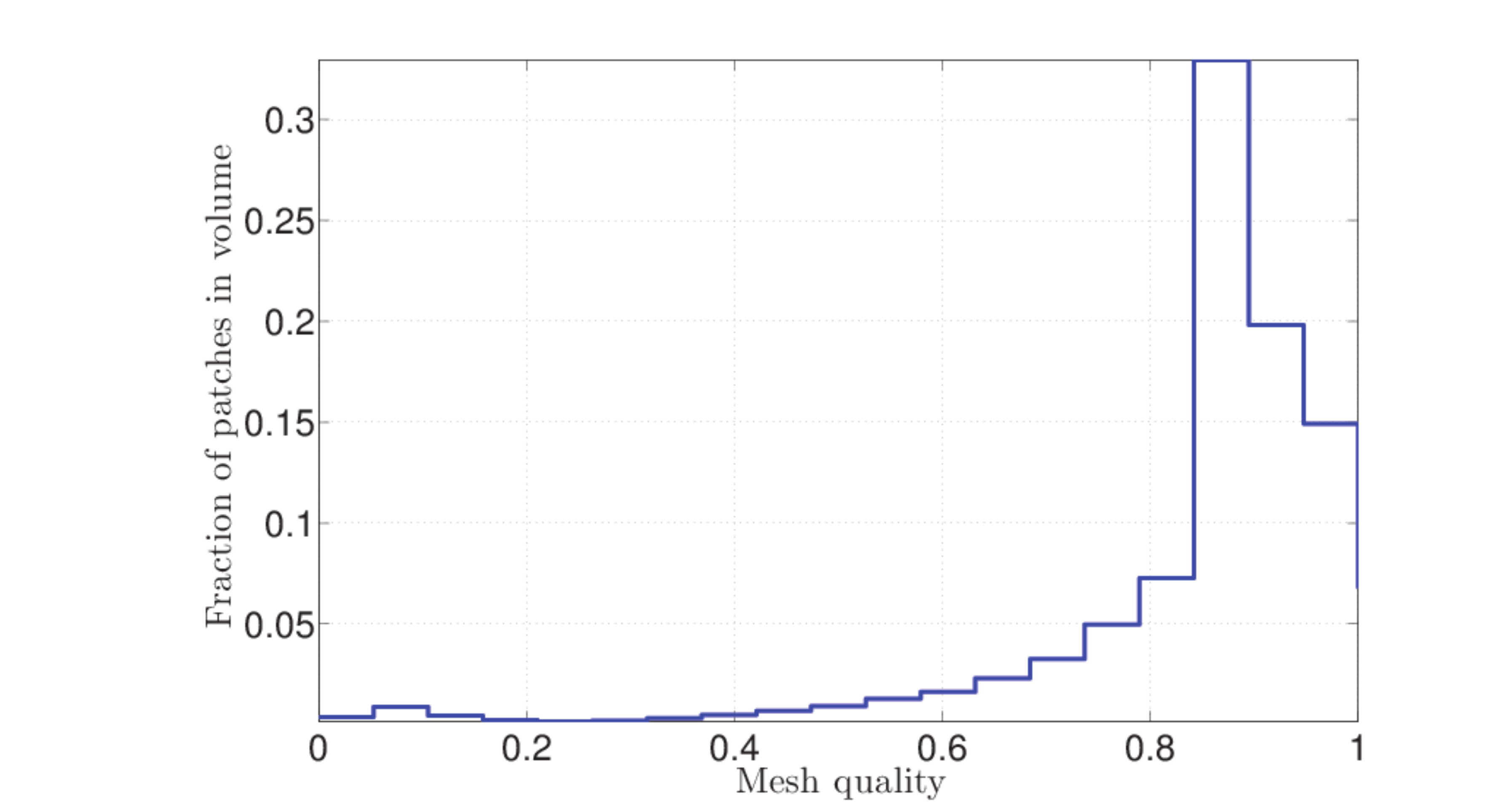}
	\caption{Calculated mesh quality over the entire 3-dimensional volume, in which $\sim 92\%$ of the patches have a quality above the rule-of-thumb value of $0.6$ for simple finite element applications~\cite{MATLAB2015}.}
	\label{fig.volmeshquality}
\end{figure}

We address the suitability of $SMOOTH$ output for finite element applications, which are predicated on the availability of surface meshes with reasonably isotropic mesh elements (equilateral, in the case of triangular) in order to avoid errors from piecewise linear interpolation. While there exist other sophisticated methods of quantifying the quality of a mesh element~\cite{Frey1999}, we implement a simple metric for triangular elements that tests their closeness to an equilateral triangle~\cite{Bank1990}. The quality of a triangle of area $A$ and side lengths $s_1$, $s_2$ and $s_3$ is defined to be:
\begin{equation}
	Q = \frac{4\sqrt{3}A}{s_1^2 + s_2^2 + s_3^2}
	\label{eq.meshquality}
\end{equation}
which gives $Q = 1$ for an equilateral triangle. Shown in Figure~\ref{fig.meshquality} are element-wise quality plots of select grains in a 3-dimensional volume. Figure~\ref{fig.volmeshquality} histograms the mesh quality over all surface elements in the entire 424 grain volume. The flattening of some of the mesh elements at the triple junctions of the 3-dimensional grains in Figures~\ref{fig.3dgrain1}, \ref{fig.3dgrainpair} and \ref{fig.3dgrainpair2} is evidenced by the slight peak in the distribution in Figure~\ref{fig.volmeshquality} at low qualities ($\sim 0.1$). 

%% file: fronttracking.tex
\begin{figure}
	\centering
	\subfloat[]{\includegraphics[width=0.25\textwidth]{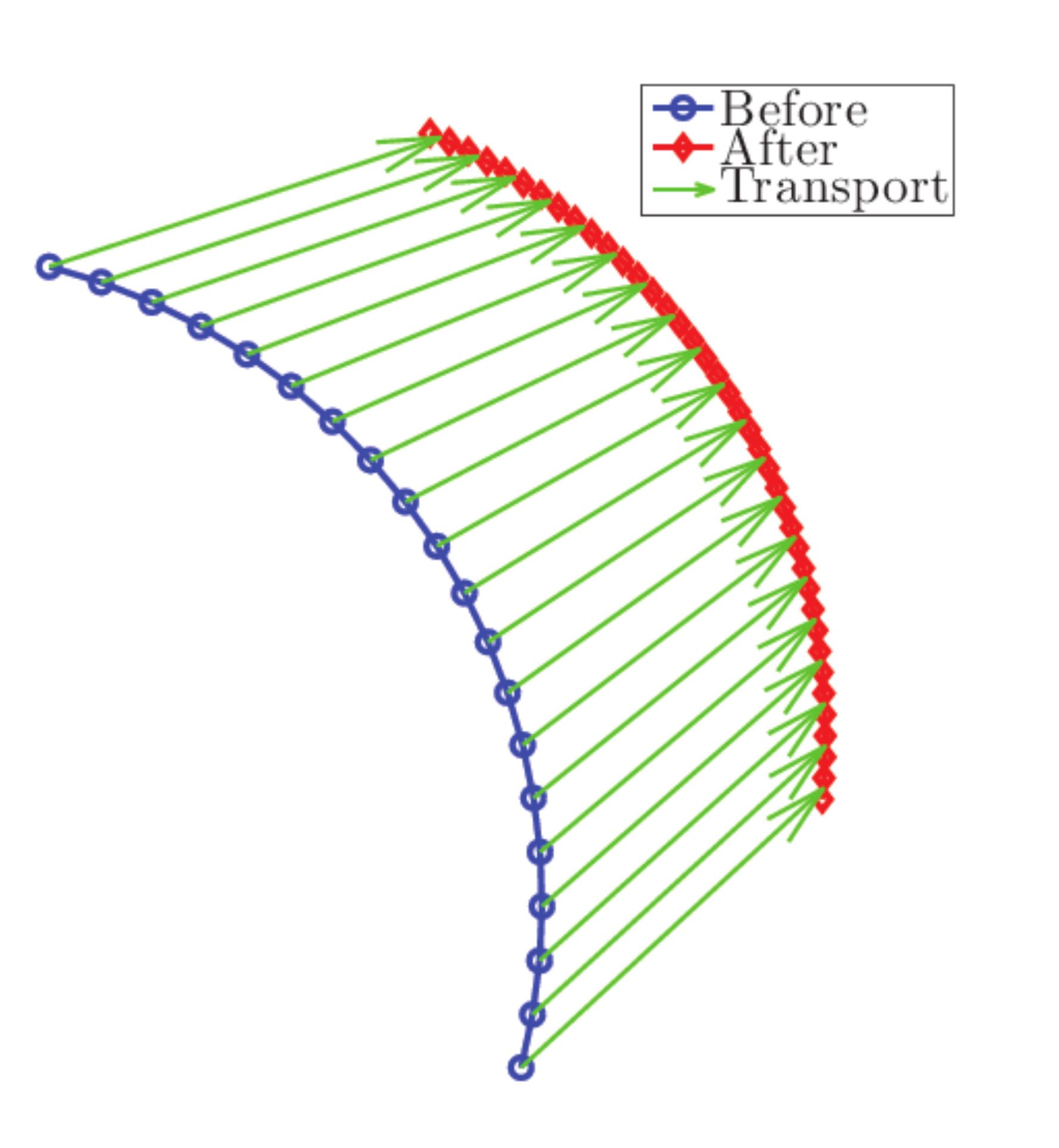}}	\hfill
	\subfloat[]{\includegraphics[width=0.25\textwidth]{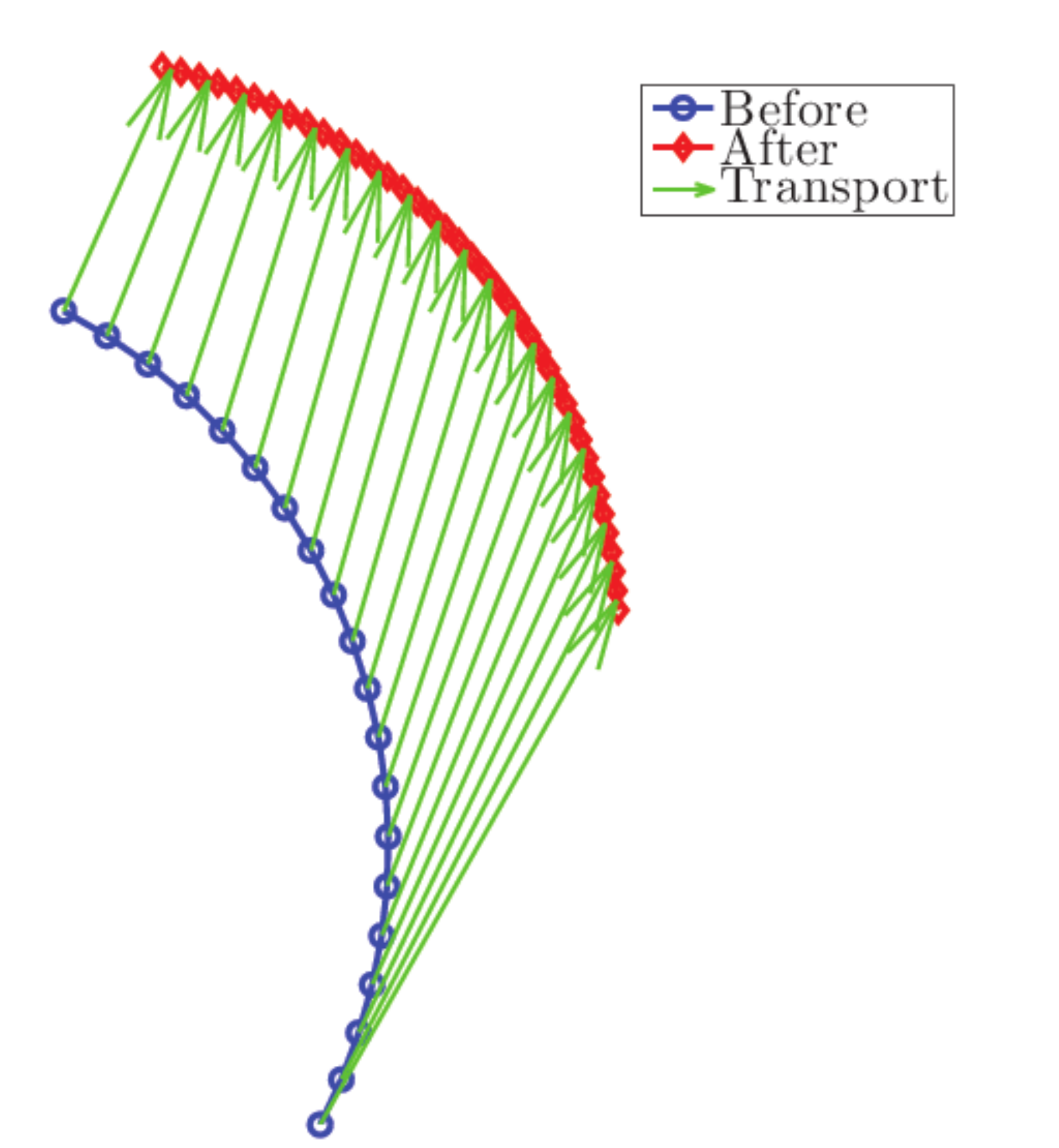}}	\hfill
	\subfloat[]{\includegraphics[width=0.25\textwidth]{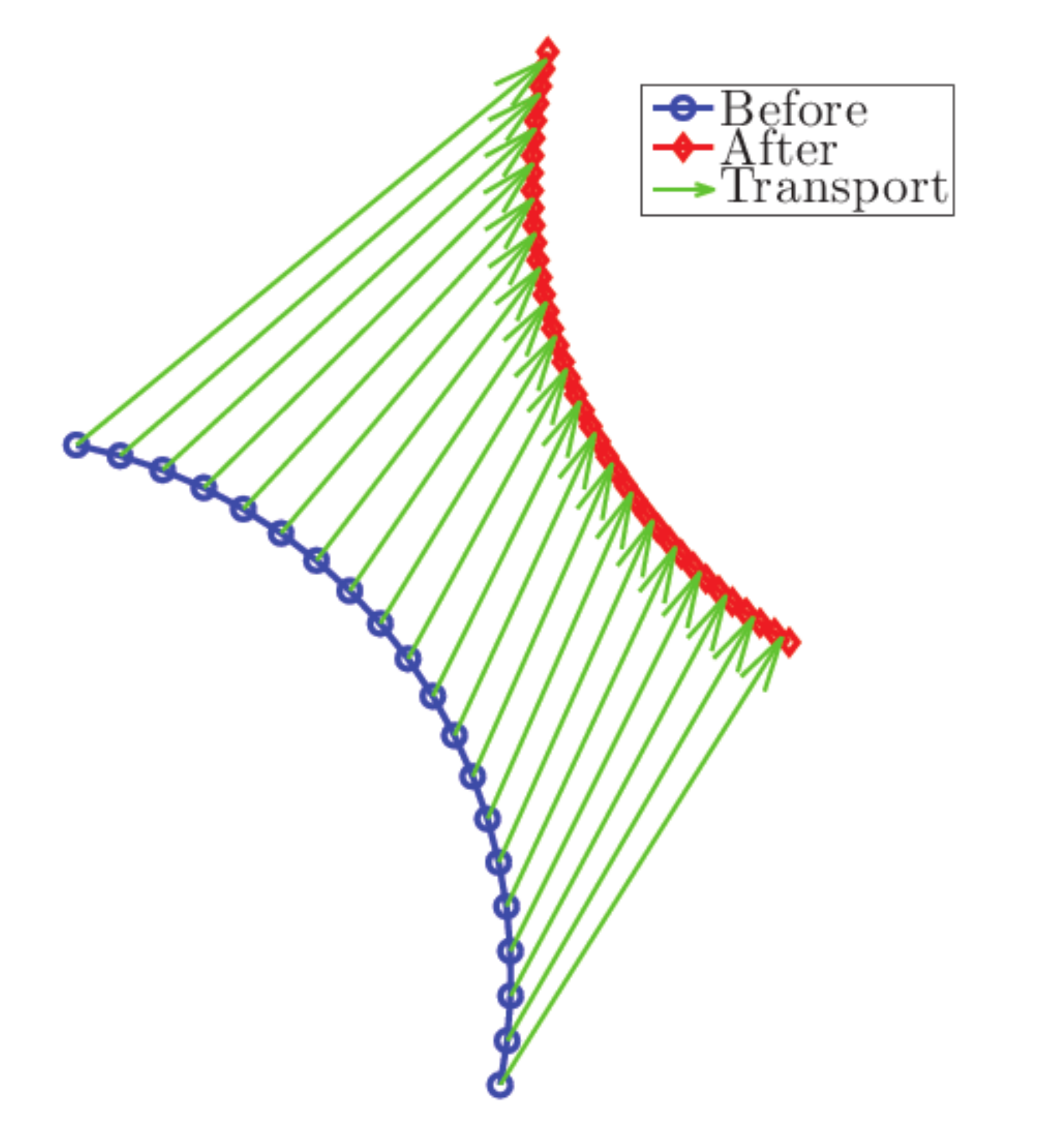}}	\hfill
	\subfloat[]{\includegraphics[width=0.25\textwidth]{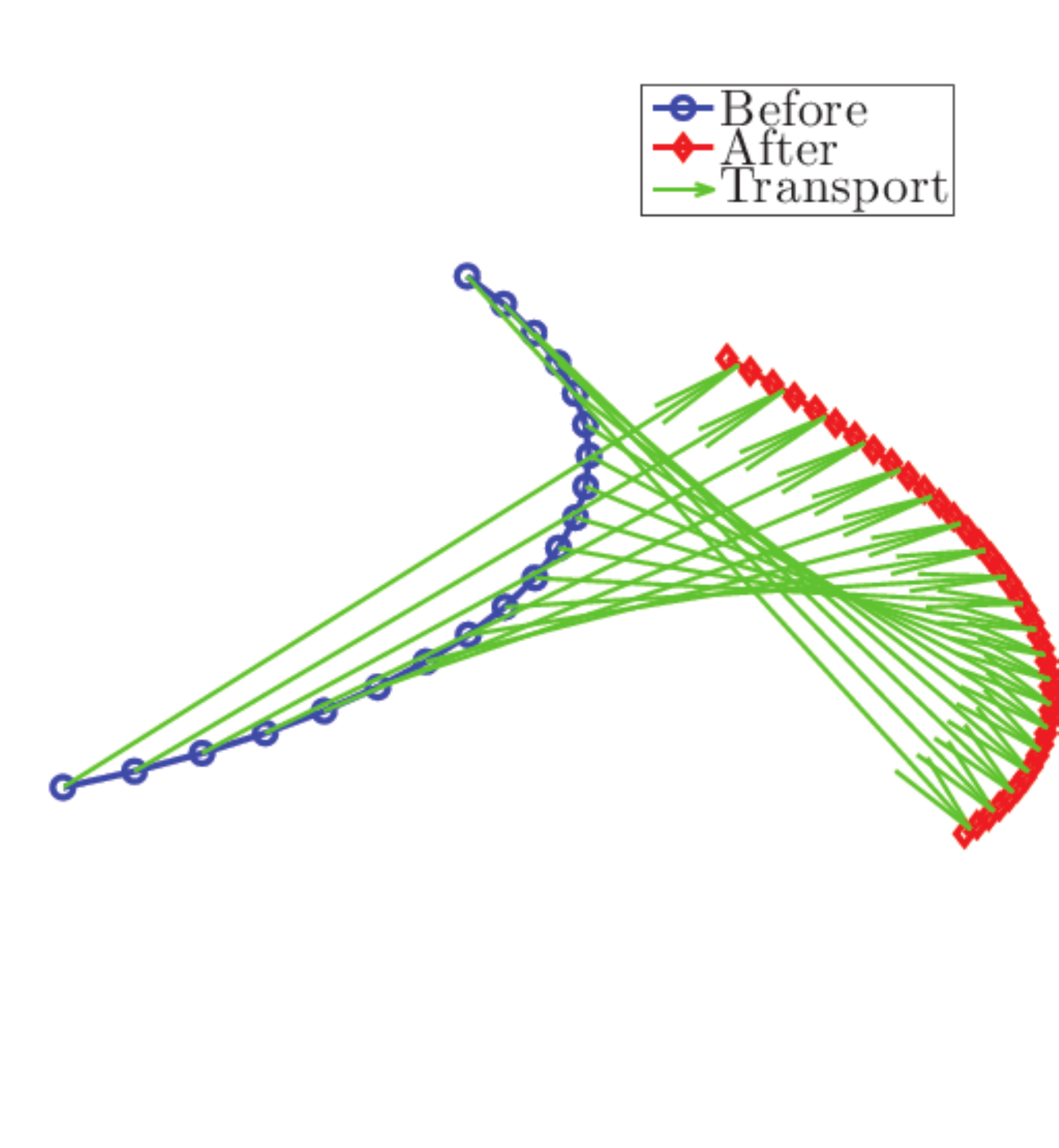}}	\hfill
	\caption{\textbf{(a)} Schematic of boundary transport illustrating the abilities of the front-tracking algorithm, additionally  with \textbf{(b)} target surface oriented differently, \textbf{(c)} inverted curvature and \textbf{(d)} 3D torsion.}
	\label{fig.OptimalTransport}
\end{figure}

%We describe in this section another potential application of the surface meshes obtained from the hierarchical surface estimation method described thus far, one that is particularly suited to modern synchrotron-based imaging techniques. The nondestructive nature of these measurements opens up an entire class of \emph{ex situ} and \emph{in situ} experiments in which bulk morphological changes can be measured, a capability not easily provided by two-dimensional destructive measurements like EBSD. This section describes a method to compute the local velocity of a grain boundary interface during grain growth, given smoothed surface meshes before and after differential migration. Such data can be obtained, for example, in an \emph{ex situ} imaging experiment. Interface velocity estimation in turn plays a role in quantifying changes in local curvature, which is known to be a driving force for interface migration. 
We describe as a final application of surface $SMOOTH$ing a methodology to infer directly the velocity field on a sampled interface on a node-to-node basis given the two snapshots of the interface before and after its differential migration, what we call the front-tracking problem. 
We present this novel computation as one riding the current wave of synchrotron-based nondestructive measurement techniques such as HEDM and DCT, and whose eventual applicability is predicated upon the estimation of the geometry of meso-scopically smooth grain boundaries.
It is generalized insofar as to account for changes in the interface geometry, such as size (or equivalently number of sample points), local curvature, surface re-orientation and torsion.
An example of of a pysical measurement that would serve as input to this algorithm would be a grain coarsening experiment by \emph{ex situ} annealing~\cite{Hefferan2010} in which bulk grain boundary structure is measured before and after a short-term annealing regimen. A satisfactory quantification of the intermediate stages of boundary migration in this manner opens up the possibility of studying the influence of local curvaure and its rate of change on coarsening. 
The emphasis on differential migration is significant because it permits us to make a qualitative assumption about the interface transport: that the sample nodes of the interface take the straightest and simplest path in the intervening space to reach the configuration described by the target set of points. A schematic of such transport is shown in Figure~\ref{fig.OptimalTransport} for a variety of morphological distortions. 

We state our problem in terms of two Cartesian point sets $\left\{\bs{x}_i\right\}$ and $\left\{\bs{y}_j\right\}$ in 3D space ($M$ and $N$ in number respectively). These two sets represent an interface before and after differential migration. Note that $M$ may not be equal to $N$. In practice this could be because the size of the grain boundary changed during transport or the sampling of the interfaces was done at different resolutions, or a combination of both. 
\begin{itemize}
	\item	We seek a set of $M$ displacement vectors $\{\Delta \bs{x}_i\}$, one for each $\bs{x}_i$ such that the set $\{\bs{x}_i + \Delta\bs{x}_i\}$ mimics the configuration of the target surface as closely as possible, with as uniform a sampling as posible.
	\item	Under these conditions, it is assumed that in the time of transport $\Delta t$ (usually obtained from experiment) of the intermediate positions of the boundary nodes are:
	\begin{align}
		\bs{x}_i^{\text{(inter)}}(\epsilon \Delta t) &= \bs{x}_i + \epsilon\Delta\bs{x}_i	\\
		\text{where } 0 &\leq \epsilon \leq 1	\notag
	\end{align}
	\item	The velocity field over the nodes is then simply given by:
	\begin{equation}
		\bs{v}_i = \frac{\Delta\bs{x}_i}{\Delta t}		\notag
	\end{equation}
	We seek to compute the node displacements $\Delta \bs{x}_i$ since the velocities $\bs{v}_i$ are estimated to first order with knowledge of the time of transport $\Delta t$. 
\end{itemize}
We demonstrate how this problem is solvable using linear optimization for which there exist a variety of numerical software packages. We calculate each displacement $\Delta\bs{x}_i$ as a weighted sum of \emph{all} possible displacements from $\bs{x}_i$ to each $\bs{y}_j$:
\begin{equation}
	\Delta \bs{x}_i = \sum_{j=1}^N C_{ij} \left(\bs{y}_j - \bs{x}_i\right)
	\label{eq.VelocityField}
\end{equation}
The unknown weights $C_{ij}$ are determined by the following linear optimization problem in the unknowns $\sigma_{ij}$:
\begin{equation}
	C_{ij} = \arg\min_{\sigma_{ij}} \left\{\sum_{i=1}^M \sum_{j=1}^N \sigma_{ij} \left|\bs{y}_j - \bs{x}_i\right|^2\right\}
	\label{eq.OptFlowObjFun}
\end{equation}
subject to the constraints:
\begin{align}
	\sigma_{ij} &\geq 0~\forall~i=1,2,\ldots,M\text{ and } j = 1, 2, \ldots,N	\label{eq.Positivity}	\\
	\sum_{j=1}^N \sigma_{ij} &= 1~\forall~i = 1, 2, \ldots, M					\label{eq.OnSurface}	\\
	\sum_{i=1}^M \sigma_{ij} &= \frac{M}{N}~\forall~j = 1, 2, \ldots, N			\label{eq.SpreadOut}
\end{align}
The objective function in Equation~\eqref{eq.OptFlowObjFun} attempts to find the smallest possible aggregate weighted squared displacement between the two point sets. The positivity constsaint~\eqref{eq.Positivity} ensures directionality of the displacements $\Delta\bs{x}_i$ towards the target surface. Constraint~\eqref{eq.OnSurface} ensures that the $\Delta\bs{x}_i$ actually terminate on the target surface and not before or after, while the equi-weight constraint~\eqref{eq.SpreadOut} adjusts the displacements so that the target surface is as uniformly sampled by the point set $\{\bs{x}_i + \Delta \bs{x}_i\}$ as can be managed. The inferred displacements of the sample nodes of a meshed surface in 3D is shown in Figure~\ref{fig.Optimal3D}. Additional examples can be seen in the supplementary material. 
\begin{figure}
	\centering
	\includegraphics[width=0.6\textwidth]{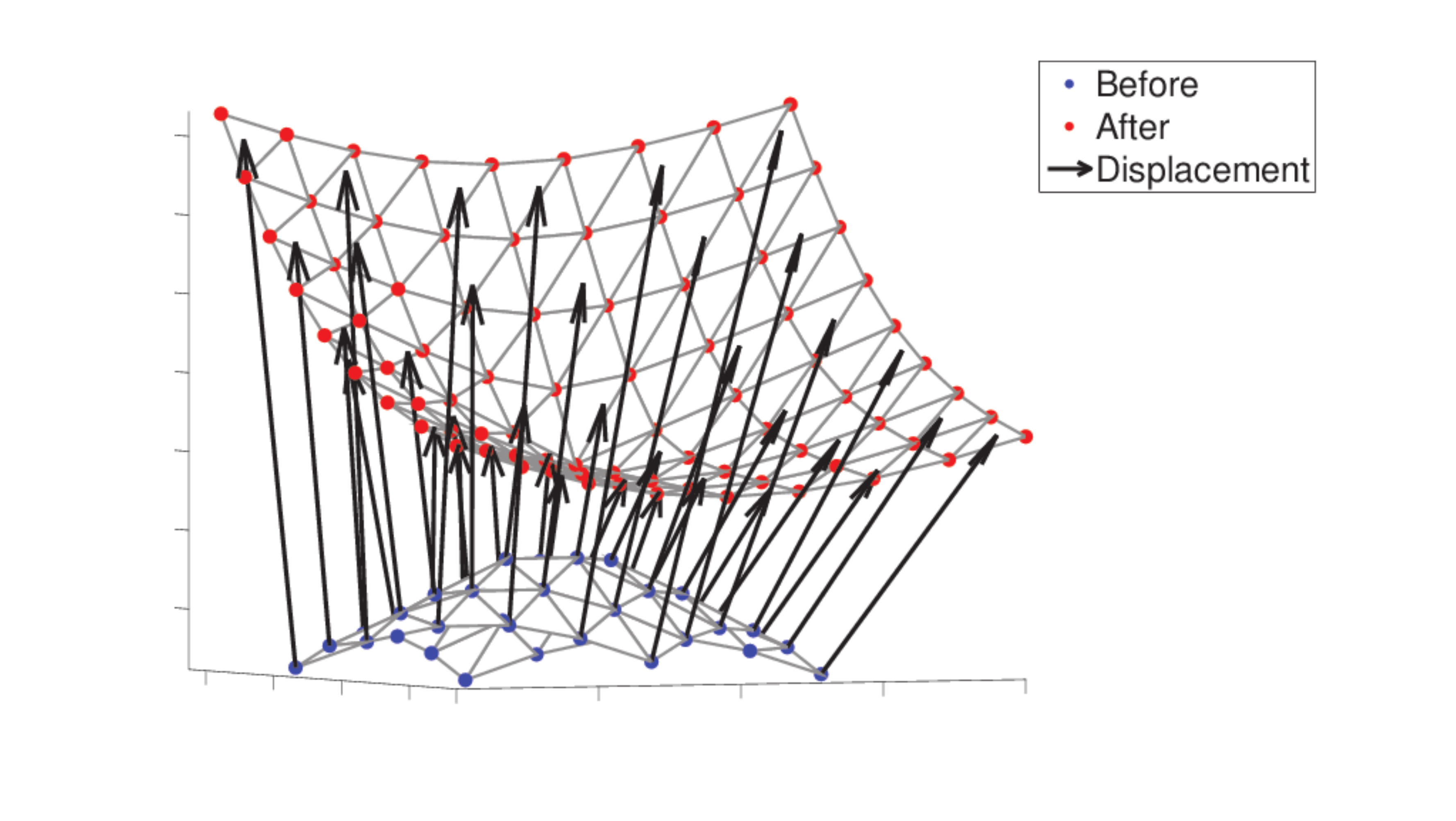}
	\caption{Inferred node transport between two interfaces with curvatures of opposite signs.}
	\label{fig.Optimal3D}
\end{figure}

It is well-known, however, that the true quantity of interest is the local \emph{normal} velocity of the interface. This is easily estimated for each surface mesh element from the inferred velocities of its constituent nodes. For example, if the nodes of a triangular surface mesh element of inclination $\hat{\bs{n}}$ have inferred displacements $\{\Delta\bs{x}_1, \Delta\bs{x}_2, \Delta\bs{x}_3\}$ then the normal displacement of the element is estimated as the projection of the displacement of the element centroid along the direction of inclination:
\begin{equation}
	\Delta\bs{x}_{\text{normal}} \simeq \frac{1}{3} \left(\Delta\bs{x}_1 + \Delta\bs{x}_2 + \Delta\bs{x}_3\right) \cdot \hat{\bs{n}}
	\label{eq.NormalVelocity}
\end{equation}
In closing, we note the following points:
\begin{itemize}
	\item	The linear optimization problem always has a unique solution and therefore the front tracking algorithm never fails. It falls on the user to decide whether the interface data satisfy the requirement of differential migration. For interface migration over extended periods of time in which the final interface is dramatically different from  the initial interface in geometry or topology, it may very well be that this configuration was not achieved through migration along straight lines in actual physical samples.
	\item	It is observed in Figure~\ref{fig.Optimal3D} that the edge nodes of the initial interface are not mapped exactly to the edges of the final interface; in fact they are mapped to a location decidedly in the interior of the target interface. This is characteristic of node transport between point sets of different sizes: $M  \neq N$. In order to retain topological consistency one may adopt a convention of hierarchical transport similar to the smooth surface estimator described earlier, in which the linear optimization algorithm is applied separately to subsets of the point sets. For example, quad points in the initial surface are explicitly transported to quad points in the target surface, triple line nodes to triple line nodes and grain boundary interior nodes to grain boundary interior nodes. The local normal displacement can be estimated as before. 
\end{itemize}

%% file: conclusion.tex
A new smoothing estimator for interface networks was demonstrated and compared to the performance of an established but generic smoothing algorithm in current use. The requisite inputs to the algorithm are:
\begin{itemize}
	\item	Discretely sampled polycrystal interface data, particularly measured by modern high-resolution synchrotron-based experiments such as nf-HEDM.
	\item	Topological characterization of the sample points as belonging to grain boundary interiors, triple lines or quad junctions. This is easily achieved by existing software packages like DREAM.3D~\cite{Groeber2014}. 
	\item	A mesh of the discretized sample points, also achieved quite simply by the QuickMesh feature of DREAM.3D.
\end{itemize}
The new algorithm organizes the topological elements of the network into an hierarchy depending on which elements physically border other elements and smooths each element set constraining its bordering elements to their fixed positions. This treatment gives the physically relevant higher-order topolological elements like triple lines and quad points their due consideration and retains the geometric discontinuities along a grain surface resulting from their existence. By its very design the smooth estimator returns a curve that passes in between the original noisy sample points, which is highly significant for pixelated images obtained from digital measurement techniques. Reorienting the sampling grid still ensures that the estimated surface lies within a pixel width of the true interface. 
Further, all elements belonging to a particular hierarchy rank in the entire volume are smoothed simultaneously so that they are ready to be used as Dirichlet boundary conditions for elements of lower rank that connect to them. The method is completely non-parametric, permitting the automated smoothing of imaged bulk structures and does not suffer from user-related effects like over-, under-smoothing or fixed-size window artifacts. Repeated applications of the smoothing on the same point set results in better smooth approximations with decreasing the extent of waviness along the smoothed surface. 

The technique is predicated on the nearest neighbor connectivity of the surface nodes being known in advance, which is easily achievable by existing algorithms and is already implemented in open-source microstructure software packages~\cite{Groeber2014}. The additional requirement of labeling surface nodes as belonging to either grain boundary interiors, triple lines or quad points is also achieved by these software packages. The case of faceted grain boundary interiors is likewise handled by appropriate labeling of this kind, for purely book-keeping purposes. This is a task external to the smoothing itself and as such is outside the scope of this work. 

The relative errors in the smoothing of known shapes are demonstrated to be a fraction of a percent for typical resolutions of microstructure imaging techniques. The estimated normals in the case of two dimensions were found to be within thresholds characteristic of bin sizes used in plots of grain boundary character distribution (GBCD). The ability to handle data of different dimensions in a generalized manner allows this technique to be used for surface experiments such as optical metallography and EBSD, as well as 3D bulk techniques like nf-HEDM. Computation of the mesh quality on a 3-dimensional grain boundary network consisting of 424 grains revealed that the overwhelming majority of mesh elements have a quality above a comfortable $0.6$ as required by simple applications. If need be one may remesh the output of the smoothing routine subject to junction constraints in order to obtain a more uniform meshing of the grain boundary interiors. 

We further detailed a new computational scheme to quantify the local displacement of a grain boundary using mathematical optimization techniques. The estimated transport of interfaces, coupled with the nonparamatric estimation of interface geometry using $SMOOTH$, potentially sets the stage for direct study of the influence of local curvature and its rate of change on interface migration in real materials, something which has not yet been achieved in three dimensions for real polycrystals.

In summary, hierarchical smoothing as a surface estimation technique, apart from yielding aesthetically pleasing results, finds applications in the computation of grain boundary character distribution (GBCD), finite-element modeling of microstructure and the study of influence of curvature on interface migration in polycrystals, the latter uniquely in conjunction with modern synchrotron-based non-destructive measurement techniques. As a further implementation-based consideration, the separate treatment of the different topological features suggests parallelizeability in order to reduce computation time.

%% file: acknowledgements.tex
The authors would like to thank David Menasche and Anthony Rollett at Carnegie Mellon University for their valuable input. This research was supported by NSF Award DMR-1105173.